\documentclass[aps,pre]{revtex4}

\usepackage{graphicx}
\usepackage{amssymb,amsmath,amssymb}
\begin{document}
\title{Stochastic thermodynamics for Ising chain and symmetric exclusion process}
\author{R. Toral}
\affiliation{IFISC (Instituto de F{\'\i}sica Interdisciplinar y Sistemas Complejos), Universitat de les Illes Balears-CSIC, 07122-Palma de Mallorca, Spain}
\author{C. Van den Broeck}
\affiliation{Stellenbosch Institute for Advanced Studies, Matieland 7602, South Africa, and\\
Hasselt University, B-3500 Hasselt, Belgium (permanent address)}
\author{D. Escaff}
\affiliation{Complex Systems Group, Facultad de Ingenier{\'\i}a y Ciencias Aplicadas, Universidad de los Andes, Avenida Monse\~nor \'Alvaro del Portillo No 12.455 Las Condes, Santiago, Chile}
\author{Katja Lindenberg}
\affiliation{Department of Chemistry and Biochemistry and BioCircuits Institute,
University of California San Diego, La Jolla, CA 92093-0340, USA}
\begin{abstract}
We verify the finite time fluctuation theorem for a linear Ising chain in contact with heat reservoirs at its ends. Analytic results are derived for a chain consisting of two spins. The system can be mapped onto a model for particle transport, namely, the symmetric exclusion process in contact with thermal and particle reservoirs. We  modify the  symmetric exclusion process to represent a thermal engine and reproduce universal features of the efficiency at maximum power. 

\end{abstract}
\maketitle
\section{Introduction}
Thermodynamics and statistical mechanics have been mutually inspiring fields of research for over 100 years. Recently, the formulation of thermodynamic laws for the description of small scale nonequilibrium systems in contact with heat and work reservoirs has deepened and extended our understanding of thermodynamics and its relation to microscopic laws.
This novel connection has been made in various different contexts, including microscopic classical and quantum descriptions, mesoscopic descriptions embodied in stochastic thermodynamics, and thermostated systems \cite{Evans:2002}-\cite{VandenBroeck:2014b}. In the present contribution, we apply stochastic thermodynamics to a prototype model of statistical mechanics, namely, a linear chain of Ising spins in contact with heat reservoirs of different temperatures at its ends. One interesting  point of our analysis is that, in contrast with most models studied so far in the context of stochastic thermodynamics, the internal dynamics of the chain is microcanonical in the sense that it is energy conserving. Nevertheless, the standard formalism of stochastic thermodynamics applies, and one of its basic predictions, the so-called fluctuation theorem, is verified. Furthermore, the system can be mapped onto a model for particle transport, namely, the symmetric exclusion process. In this respect, we note that, with a proper interpretation of the boundary conditions, the system can function as a small scale thermal engine. We verify another prediction of stochastic thermodynamics, the universality of efficiency at maximum power.

The outline of the paper is as follows. In section \ref{sec:linear} we define the model and its different interpretations in terms of energy and particle transport, and we review the relation with the symmetric exclusion process. In section \ref{sec:efficiency} we discuss its use as a heat engine and compute its efficiency at maximum power, showing that it displays some universal features. In section \ref{sec:entropy} we discuss and numerically check the validity of the fluctuation theorem. In section \ref{sec:large} we analytically derive the large deviation function for the case of two spins. Finally, in section \ref{sec:discussion} we summarize the main conclusions. The more technical details of the paper are presented in the appendices. 

\section{Linear Ising chain and symmetric exclusion process}
\label{sec:linear}
We consider a $1$-d Ising chain with $M$ nodes and nearest neighbor interactions. To each configuration $\{s\}=(s_1,\dots,s_M)$, $s_i=\pm 1$, we assign the value of the Hamiltonian function
\begin{equation}
{\cal H}(\{s\})=-\frac{\epsilon}{2} \sum_{i=1}^{M-1}s_is_{i+1}.
\end{equation} 
This can also be written as ${\cal H}=\frac{\epsilon}{2}\left[\ell_1+\ell_2+\dots+\ell_{M-1}\right]$, where $\ell_i=-s_is_{i+1}$ is a variable associated with the link between spins $i$ and $i+1$. 
 As for boundary conditions, we consider the situation in which $s_1$ is in contact with a heat reservoir $\mathbf{B_1}$ at temperature $T_1$ and $s_M$ with another heat reservoir $\mathbf{B_2}$ at temperature $T_2>T_1$. Energy is transferred in the form of heat from $\mathbf{B_2}$ to $\mathbf{B_1}$. 
The connections to the reservoirs induce a stochastic dynamics in which spins $s_1$ and $s_M$ update their states using heat-bath canonical rates at temperatures $T_1$ and $T_2$, respectively. More precisely, the probabilities that the spins $s_1$ and $s_M$ adopt particular values are given by
\begin{equation}
\text{prob}(s_1)=\frac{1}{1+e^{-\epsilon s_1 s_2/kT_1}}\mbox{~~~~~~prob}(s_M)=\frac{1}{1+e^{-\epsilon s_{M-1}s_M/kT_2}},
\end{equation}
where $k$ is the Boltzmann constant. The dynamics of the internal spins $s_i,\, i=2,\dots,M-1$ are assumed to be microcanonical in the sense that a spin can change its state $s_i\to -s_i$ 
provided that energy is conserved. In other words, spin $s_i$ can only flip provided that its neighbors are in opposite states, $s_{i-1}+s_{i+1}=0$.

These updating rules induce Markovian dynamics between the different configurations with rates $\omega(\{s\}\to\{s'\})$ which are different from zero only if configuration $\{s'\}$ differs from configuration $\{s\}$ in the value of a single spin. 
The probability $P(\{s\},t)$ for a configuration $\{s\}$ at time $t$ thus satisfies the following master equation:
\begin{equation}
\label{me}
\frac{dP(\{s\},t)}{dt}=\sum_{\{s'\}}\left[\omega(\{s'\}\to\{s\})P(\{s'\},t)-\omega(\{s\}\to\{s'\})P(\{s\},t)\right].
\end{equation}
We refer to Appendix I for a detailed description of the numerical procedure which we use to obtain the statistical thermodynamic properties for this process.

We are interested in this Ising model with a finite number of spins as a small scale nonequilibrium system, for which stochastic thermodynamics can be applied, see e.g. \cite{VandenBroeck:2014b} for a simple introduction. This formalism can be applied without modification for stochastic systems with ``internal" transitions, provided they satisfy detailed balance with respect to a microcanonical distribution rather than the canonical or grand-canonical distribution that apply to rates describing the contact with the reservoirs. Hence, only transitions between states of the same energy are possible and, since  the corresponding microcanonical probabilities are equal, these rates are equal. With this proviso, we will verify and discuss the stochastic heat transport and corresponding stochastic entropy production in the Ising chain. 

\begin{figure}
\includegraphics[width=16cm,angle=0]{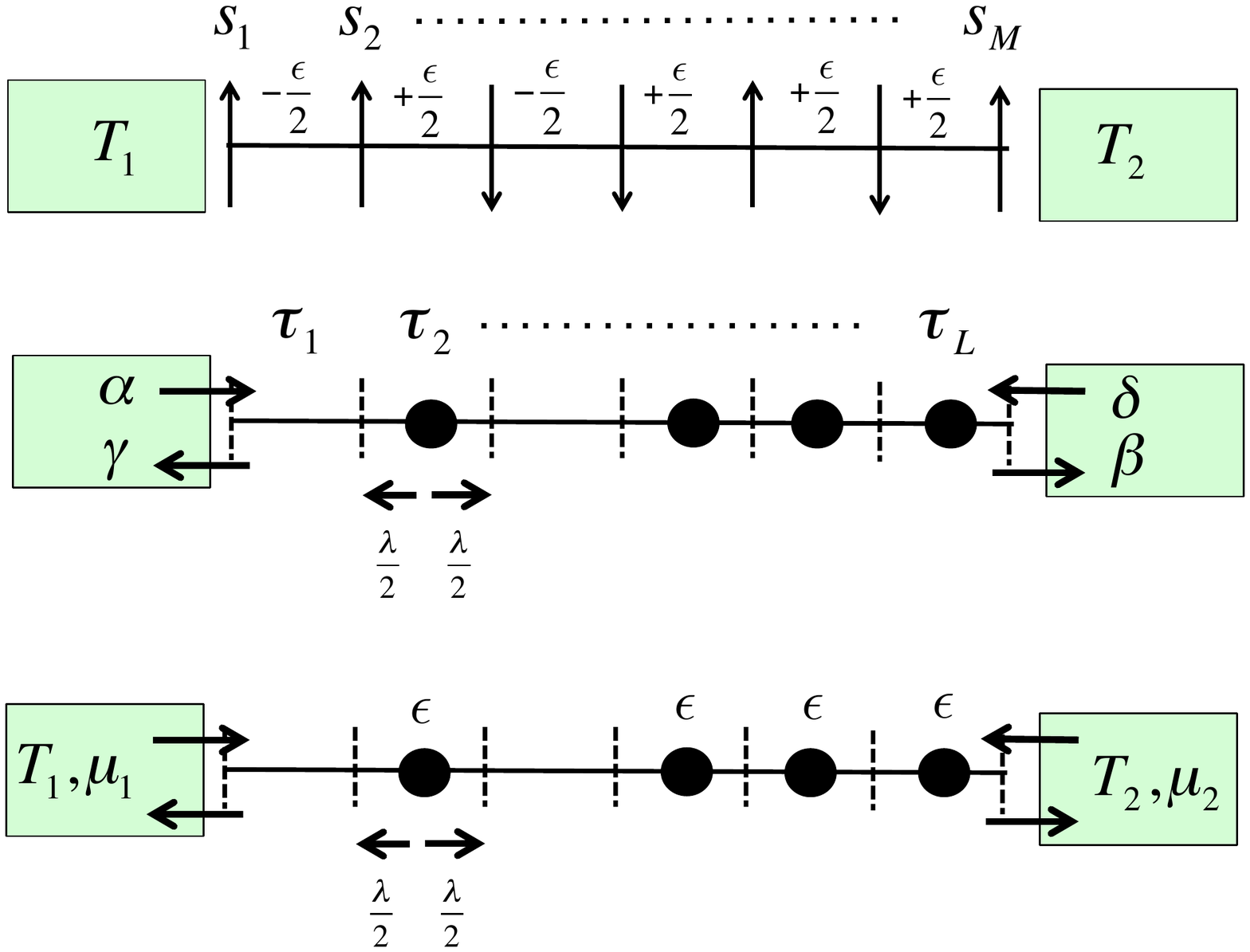}
\caption{\label{Fig:scheme}Schematic representation of the three different interpretations of our model. Top row corresponds to a ``standard" Ising chain with an energy flux between two heat reservoirs at different temperatures $T_1$ and $T_2$. The middle row represents particle transport, with  insertion/removal rates $\alpha,\,\beta,\,\delta,\,\gamma$ at the ends. The lower row features a thermal engine with both energy and particle transport between two heat and particle reservoirs at respective temperatures $T_1,\,T_2$ and chemical potentials $\mu_1,\,\mu_2$. }
\end{figure}

The above model is known to be isomorphic to one for particle transport, namely, the one-dimensional symmetric simple exclusion process. This model has been studied intensely in the past decades. It is one of the rare instances for which the exact expression for the stationary (nonequilibrium) distribution $P_\textrm{st}(\{s\})$ has been derived \cite{Derrida:2011}. The mapping of the Ising version to the particle version is as follows (see Fig.~\ref{Fig:scheme}): To each configuration $(s_1,\dots,s_M)$ of the Ising chain we assign a configuration $\tau_1,\dots,\tau_{L}$ with $L=M-1$ and $\tau_i=\frac12(1-s_is_{i+1})=\frac12(1+\ell_i)$ such that $\tau_i=1\,(\textrm{resp. } 0)$ if the energy of the link is $\ell_i=+1\,(\textrm{resp. }-1)$. We interpret $\tau_i=1\,(\textrm{resp. }0)$ as the presence (resp. absence) of a particle in the link between nodes $i$ and $i+1$. In the exclusion process particles are introduced on the site $i=1$ at a rate $\alpha$ only if another particle does not occupy this site; a particle on site $1$ can be removed with rate $\gamma$; a particle can be introduced on site $L$ with rate $\delta$, provided the site is not already occupied; and, finally, a particle on site $L$ can be removed with rate $\beta$. Particles inside the chain can move right or left with a rate $\lambda$ (setting the unit of time) only if the site to which the particle wants to jump is not occupied. There are $2^M$ configurations $\{s\}=(s_1,\dots,s_M)$ and $2^{L}$ configurations $\{\tau\}=(\tau_1,\dots,\tau_{L})$. A configuration $(\tau_1,\dots,\tau_{L})$ is equivalent to two configurations $(s_1,\dots,s_M)$ which differ only in a global spin flip. If $\{ \tau\}$ and $\{s\}$ are two equivalent configurations, then $P(\{\tau\})=2P(\{s\})$. The stochastic dynamics of the two versions (Ising and particle) of the model are also equivalent if the insertion and removal rates are related to the temperatures by
\begin{eqnarray}\label{eq:equivalence}
\alpha=\lambda p_1,\,\gamma=\lambda (1-p_1),\,\delta=\lambda p_2,\,\beta=\lambda (1-p_2),
\end{eqnarray}
with $\lambda$ the time-scale factor between the two models, and where we have defined
\begin{equation}
\label{psfp}
p_{1}=\dfrac{1}{1+e^{\epsilon/kT_{1}}}  \mbox{~~~~~~~~~~}p_{2}=\dfrac{1}{1+e^{\epsilon/kT_{2}}}.
\end{equation}

\begin{figure}
\includegraphics[width=8cm,angle=0]{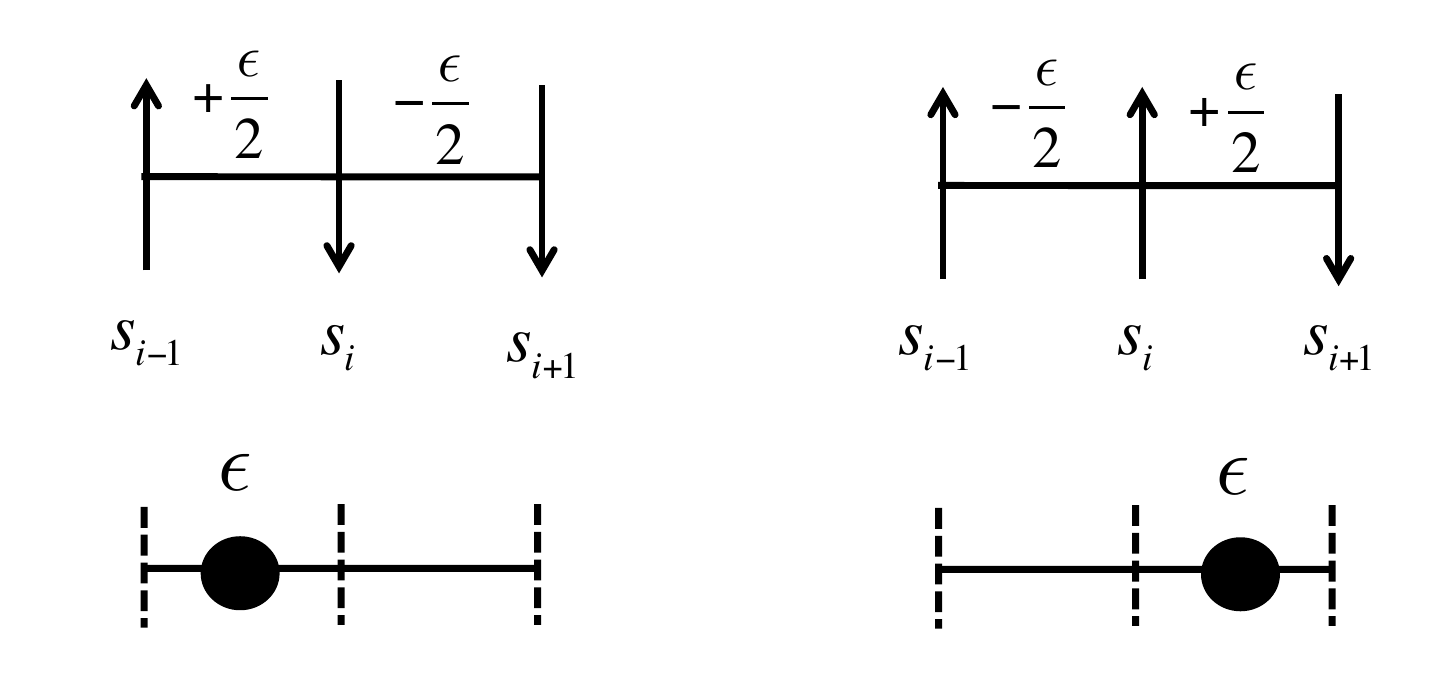}
\caption{\label{Fig:spinflip}Mapping between the spin and the particle interpretations. When the spin $s_i$ is flipped, an amount of energy $\epsilon$ is moved along the chain. This can also be interpreted as the movement of a particle carrying an energy $\epsilon$.}
\end{figure}

Our interest in this connection has a different focus:
as the reservoirs can be understood to specify both temperature and chemical potential, the system can operate as a small scale Carnot engine. Its corresponding properties can again be studied from the point of view of stochastic thermodynamics. The model is closely related to that of particle-energy transport considered in \cite{Esposito:2009b,VandenBroeck2012,Harbola2014}, with the difference that two particles can not occupy the same site, see also \cite{becker}. To make the connection with a thermal engine we include, in addition to the above prescription, a new ingredient such that particle motion implies both a particle and an energy flux. This is simply achieved by identifying the presence of a particle with the presence of an energy amount $\epsilon$. To give a concrete example, one can imagine that the particle sites correspond to quantum dots and that the appearance of a particle corresponds to an excitation in the quantum dot from energy zero to energy $\epsilon$.
When a particle moves from a site to a neighboring site, this energy is moved accordingly. Note that this is in fact also taking place in the corresponding spin system, as, for example, a spin-up flip of a spin-down between a spin-up and spin-down neighbor, corresponds to a change of the spin-pair energies from $\epsilon/2,-\epsilon/2$ to $-\epsilon/2,\epsilon/2$. Hence an amount of energy equal to $\epsilon$ has moved along the spin chain, see Fig.~\ref{Fig:spinflip}. The consideration of the energy associated with the presence of a particle becomes particularly interesting if we describe the contact with the reservoirs as an exchange with a particle and heat reservoir, say with respective temperatures and chemical potentials $T_1<T_2$ and $\mu_1>\mu_2$. Hence when a particle enters from reservoir $1$, the required energy $\epsilon$ is provided by a chemical work contribution $\mu_1$ plus an extra contribution $\epsilon-\mu_1$ which is heat provided by the same reservoir. 
To properly describe the exchange with the reservoirs, the insertion probabilities now have to obey the grand-canonical rule, denoted by a prime to distinguish them from the canonical situation, cf. Eq.~(\ref{psfp}):
\begin{eqnarray}
p_1'=\frac{1}{1+e^{(\epsilon-\mu_1)/kT_1}}   \mbox{~~~~~~~~~~}  p_2'=\frac{1}{1+e^{(\epsilon-\mu_2)/kT_2}}.
\label{insertion1}
\end{eqnarray}

\section{Efficiency}
\label{sec:efficiency}
We first focus on the Ising spin chain version and discuss the heat transport through this system. In a finite time $t$ an amount of heat $Q_i(t),\,i=1,2$ will be extracted from the reservoir $\mathbf{B_i}$. In the long time limit, a steady state regime is reached in which the cumulative average heats increase linearly with $t$, corresponding to a time-independent heat current $J_Q$,
\begin{equation}
\label{jt}
J_Q=\frac{\langle Q_2(t)\rangle }{t}=-\frac{\langle Q_1(t)\rangle }{t}>0.
\end{equation}
Interpreted within the framework of the model for particle transport, the heat current $J_Q$ is related to the particle current $J$ by $\epsilon J=J_Q$, a property that has been called strong coupling~\cite{VandenBroeck:2005}. In 
Appendix II, we obtain the following exact expression for this net current (for any value of the number of spins $M$):
\begin{equation}\label{hc1}
J=\frac{p_2-p_1}{M}. 
\end{equation}
Introducing Eq.~(\ref{psfp}), the corresponding expression of the heat current is
\begin{equation}\label{hc2}
J_Q=\dfrac{\tanh(\epsilon/2kT_1)-\tanh(\epsilon/2kT_2)}{2M}\epsilon.
\end{equation}

We next turn to the interpretation of the model as a particle transport symmetric exclusion process in contact with heat and particle reservoirs with respective temperatures and chemical potentials $T_1,\,T_2$ and $\mu_1,\,\mu_2$. In this interpretation, the system now transports both heat and chemical energy. With this interpretation the symmetric exclusion process can function as a thermal machine where a heat flow from high to low temperature drives a particle flow (hence a production of work) from low to high chemical potential. It is thus possible to calculate the efficiency of this engine and to verify its expected universal properties. When a particle hops, it takes with it a given amount of energy. As  physical realizations of such a situation, we cite the hopping of an electron in a linear array of quantum dots or of an excitation in a linear array of states (for example, a linear polymer).

As discussed above, when a particle is removed from reservoir $\mathbf{B_1}$ with chemical potential $\mu_i$, the corresponding energy flow $J_{Q_i}$ contains a chemical work component. More precisely, we have:
 \begin{eqnarray}
 J_{Q_i}&=&(\epsilon-\mu_i)J,\quad i=1,2.
 \end{eqnarray}
 At the steady state, the particle current $J$ can be copied from Eq.~(\ref{hc1}) using appropriate insertion rates Eqs.~(\ref{insertion1}):
\begin{equation}
J=\frac{p_2'-p_1'}{M}=\dfrac{\tanh(x_1/2)-\tanh(x_2/2)}{2M},
\end{equation}
where we have defined $x_i=\dfrac{\epsilon-\mu_i}{kT_i}$.

The transport of particles from the high temperature low chemical potential reservoir $\mathbf{B_2}$ to the one with low temperature and high chemical potential $\mathbf{B_1}$ is tantamount to a chemical engine. The power (chemical energy produced per unit of time) is given by
\begin{equation}
{\cal P}=(\mu_1-\mu_2)J=kT_2[x_2-(1-\eta_C)x_1]J
\end{equation}
and the corresponding efficiency reads
\begin{equation}
\label{etan}
\eta=\frac{\cal P}{J_{Q_2}}=1-(1-\eta_C)\frac{x_1}{x_2},
\end{equation}
with $J_{Q_2}$ the heat flow out of the hot reservoir $1$ and $\eta_C=1-T_1/T_2$ the Carnot efficiency.

To compute the efficiency $\eta^*$ at maximum power we search for the values $x_1^*$ and $x_2^*$ that maximize the power:
\begin{equation}
\left.\frac{\partial {\cal P}}{\partial x_1}\right|_{(x_1^*,x_2^*)}=
\left.\frac{\partial {\cal P}}{\partial x_2}\right|_{(x_1^*,x_2^*)}=0.
\end{equation}
These equations determining $(x_1^*,x_2^*)$ are transcendental. A recursive solution can be found using a series expansion in $\eta_C$:
\begin{eqnarray}
x_1^*&=&a_0+a_1\eta_C+a_2\eta_C^2+a_3\eta_C^3+\dots\label{x1a0}\\
x_2^*&=&b_0+b_1\eta_C+b_2\eta_C^2+b_3\eta_C^3+\dots\label{x2b0}
\end{eqnarray}
As the case $\eta_C=0$ is degenerate (the extrema of ${\cal P}$ are then achieved by any $x_1^*=x_2^*$), the exact calculation of the expansion is somewhat tricky, cf. Appendix III for details. The result is (the numerical coefficients are given to six significant digits):
\begin{equation}
\label{etaexp}
\eta^*=\frac12\eta_C+\frac18\eta_C^2+0.0774919\eta_C^3+0.0540545\eta_C^4+0.0396952\eta_C^5+0.0301064\eta_C^6+O(\eta_C^7)
\end{equation}
Note that first two coefficients have the universal value predicted in \cite{Esposito:2009b}.

\section{Entropy production: Fluctuation theorem}
\label{sec:entropy}
The fluctuation theorem is one of the most spectacular recent results in statistical mechanics \cite{Evans:2002}-\cite{VandenBroeck:2014b}. It was originally discovered in thermostated systems in its time-asymptotic form, and mathematically linked to a symmetry property of the largest eigenvalue of a tilted evolution operator. Later on, it was realized that various versions of the fluctuation theorem can be derived, some of which are valid also at all times.
The asymptotic form of the fluctuation theorem has been studied in some detail in the asymmetric exclusion process. Our intention here is to study the finite time version. 
Stochastic thermodynamics predicts (in the absence of time-dependent driving such as considered in the Ising chain problem) that the probability $P(\Delta S)$ of observing a {\bf total} entropy change $\Delta S$ during a given (finite) time interval $t$ is exponentially larger than the probability for observing a corresponding decrease, 
\begin{equation}\label{FT}
\frac{P(\Delta S)}{P(-\Delta S)}=e^{\Delta S/k}.
\end{equation}

It is important to realize that, in order to verify this prediction, we need to evaluate the {\bf total} entropy change $\Delta S$.
The entropy change in the reservoirs, which is (the stochastic amount of) heat over temperature for each of the reservoirs, dominates the time-asymptotic limit, as it grows without bound with time (and in fact on average proportional to time). But at finite times, one needs to also measure the (bounded) stochastic entropy change of the system. This is a much more intricate quantity.  One essential point in stochastic  thermodynamics is that one can define the (stochastic nonequilibrium) entropy of a given micro state $\{s\}$ in terms of the probability $P(\{s\})$ for this state by $S_\textrm{system}=-k\log P(\{s\})$. For simplicity, we will operate under steady state conditions, so that we only need to determine the steady state probability $P_\textrm{st}$. We have already indicated that the Ising chain/symmetric exclusion process is one of the very few instances in non-equilibrium statistical mechanics for which an exact expression for the stationary distribution $P_\textrm{st}(\{s\})$ has been derived. Unfortunately, the exact expression only becomes explicit in the limit of a large system, far beyond the sizes for which we would like to verify the finite-time fluctuation theorem. Hence we have resorted to complimentary methods -one algebraic, one numerical- to calculate $P_\textrm{st}(\{s\})$ essentially exactly for the small systems of interest, see Appendix IV for more details.  

\begin{figure}
\includegraphics[width=8cm,angle=0]{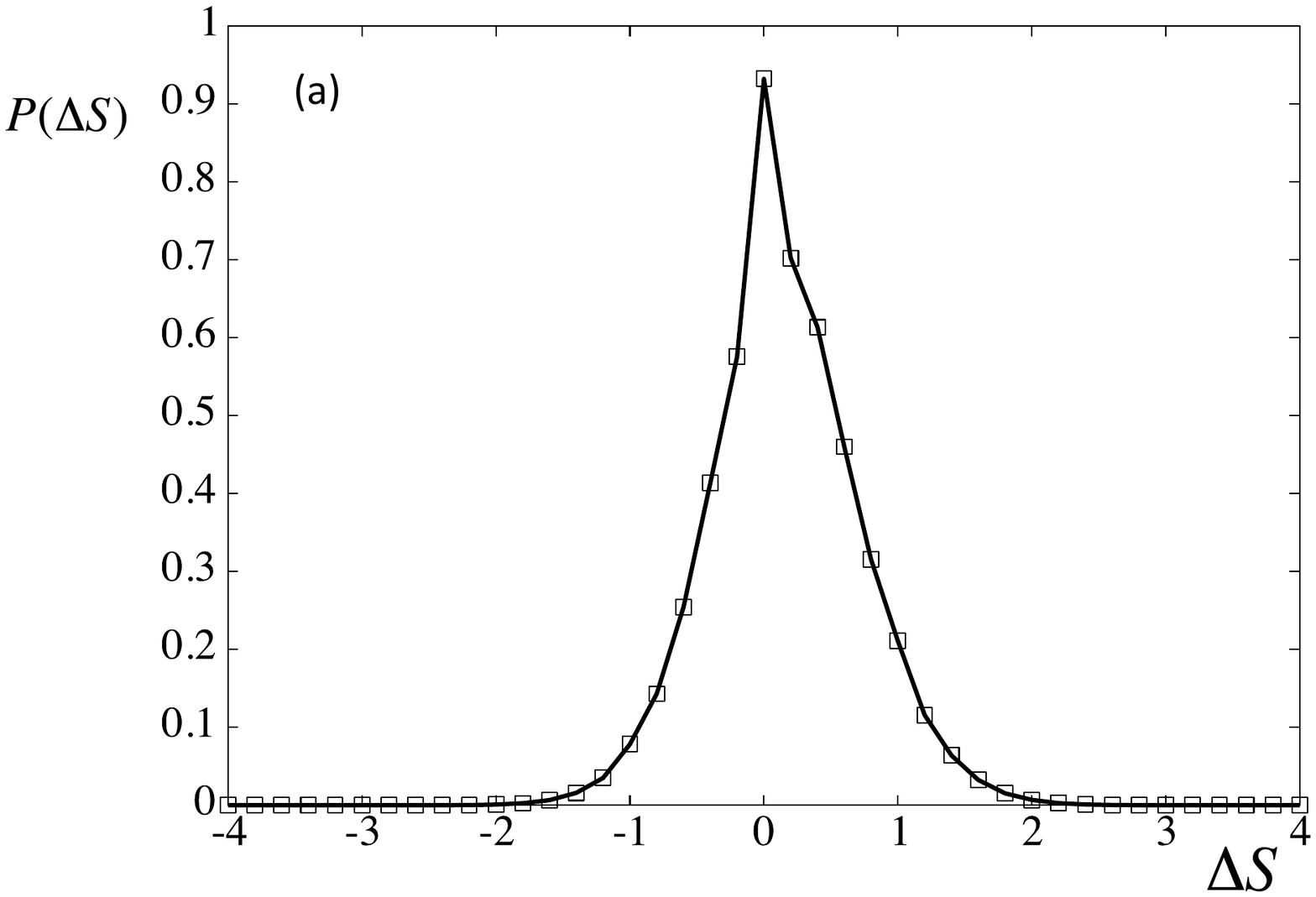}
\includegraphics[width=8cm,angle=0]{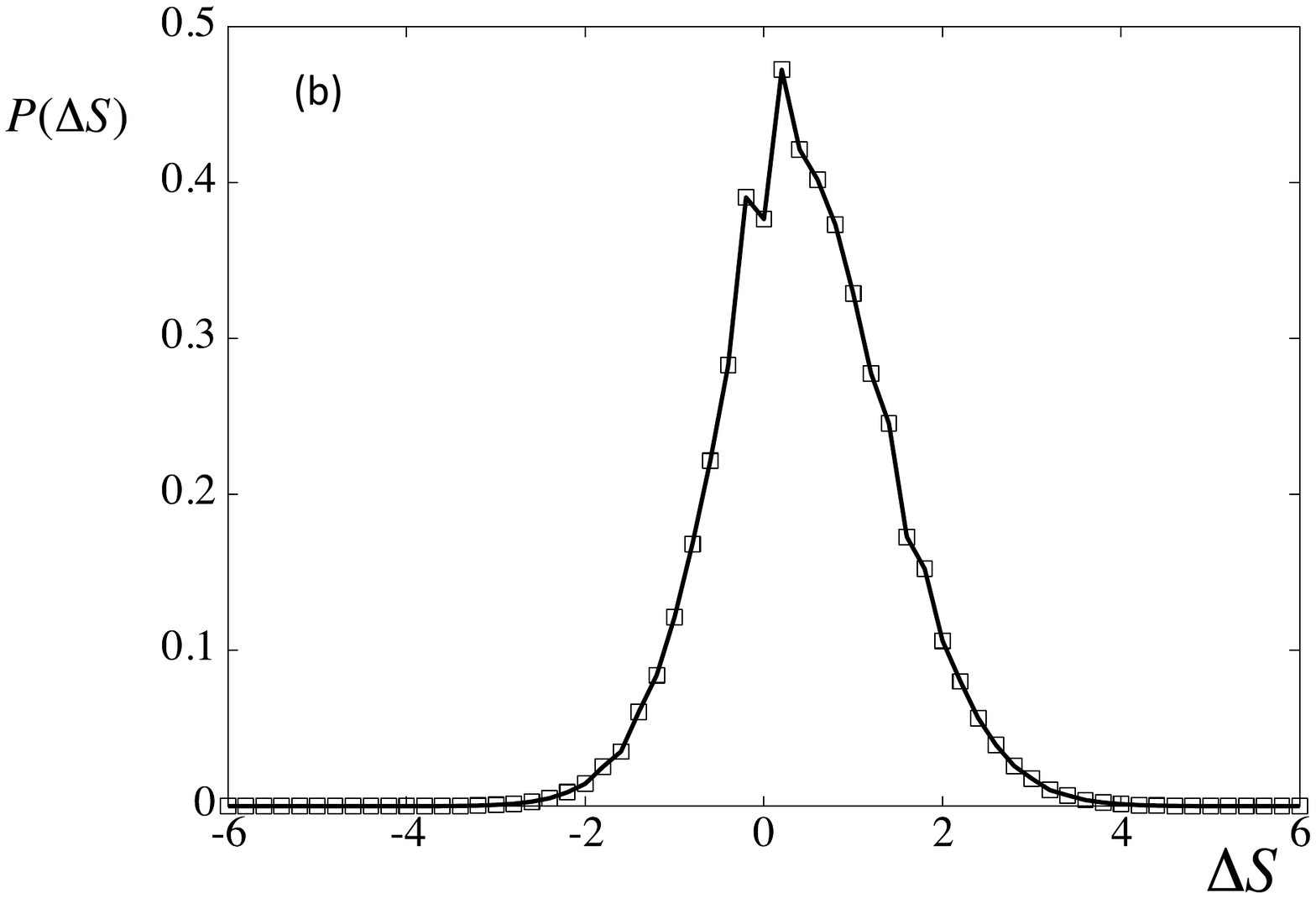}
\includegraphics[width=8cm,angle=0]{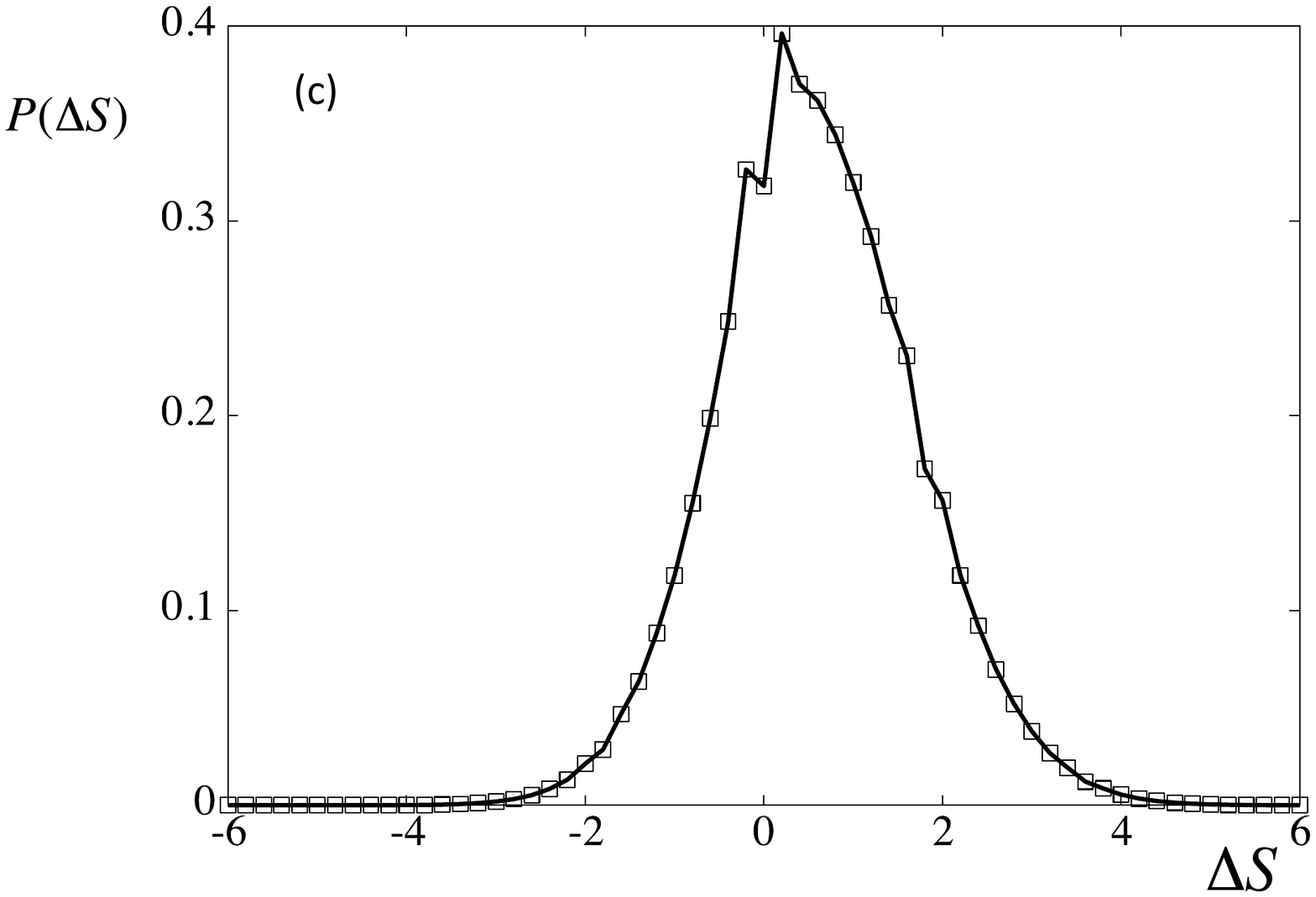}
\includegraphics[width=8cm,angle=0]{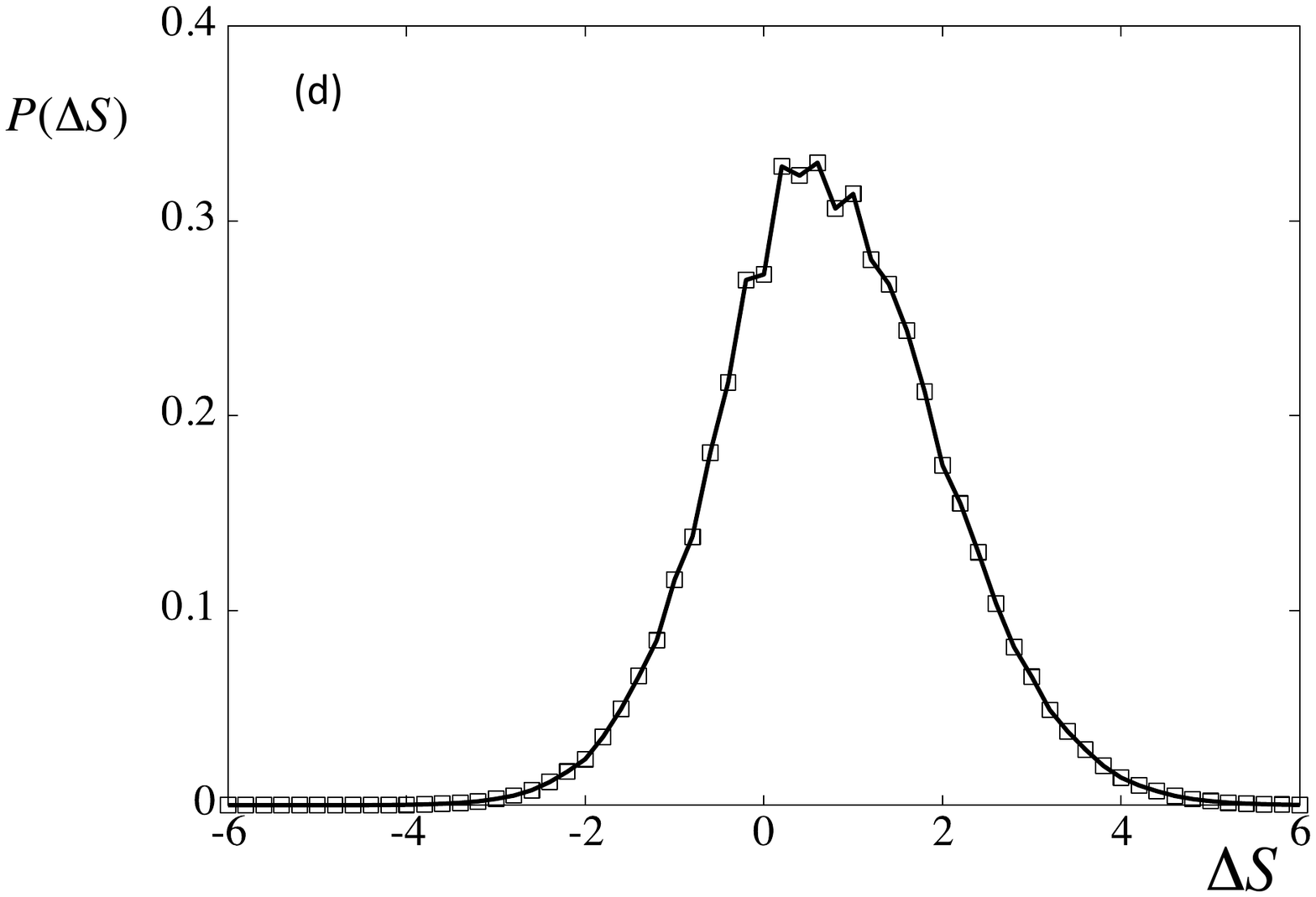}
\caption{\label{Fig:ps}$P(\Delta S)$, probability of finding a value of the entropy increase $\Delta S$, after a time of $t=10$ Monte Carlo steps, with $T_1=1$ and:  $T_2=2$, panel (a); $T_2=5$, panel (b); $T_2=10$, panel (c); $T_2=\infty$, panel (d). Results obtained by an average over $K=10^9$ configurations. The number of spins is $M=10$. A bin size of $\Delta S=0.2$ has been used in the construction of the histogram. We set $\epsilon=2,\,k=1$.
}
\end{figure}

\begin{figure}
\includegraphics[width=8cm,angle=0]{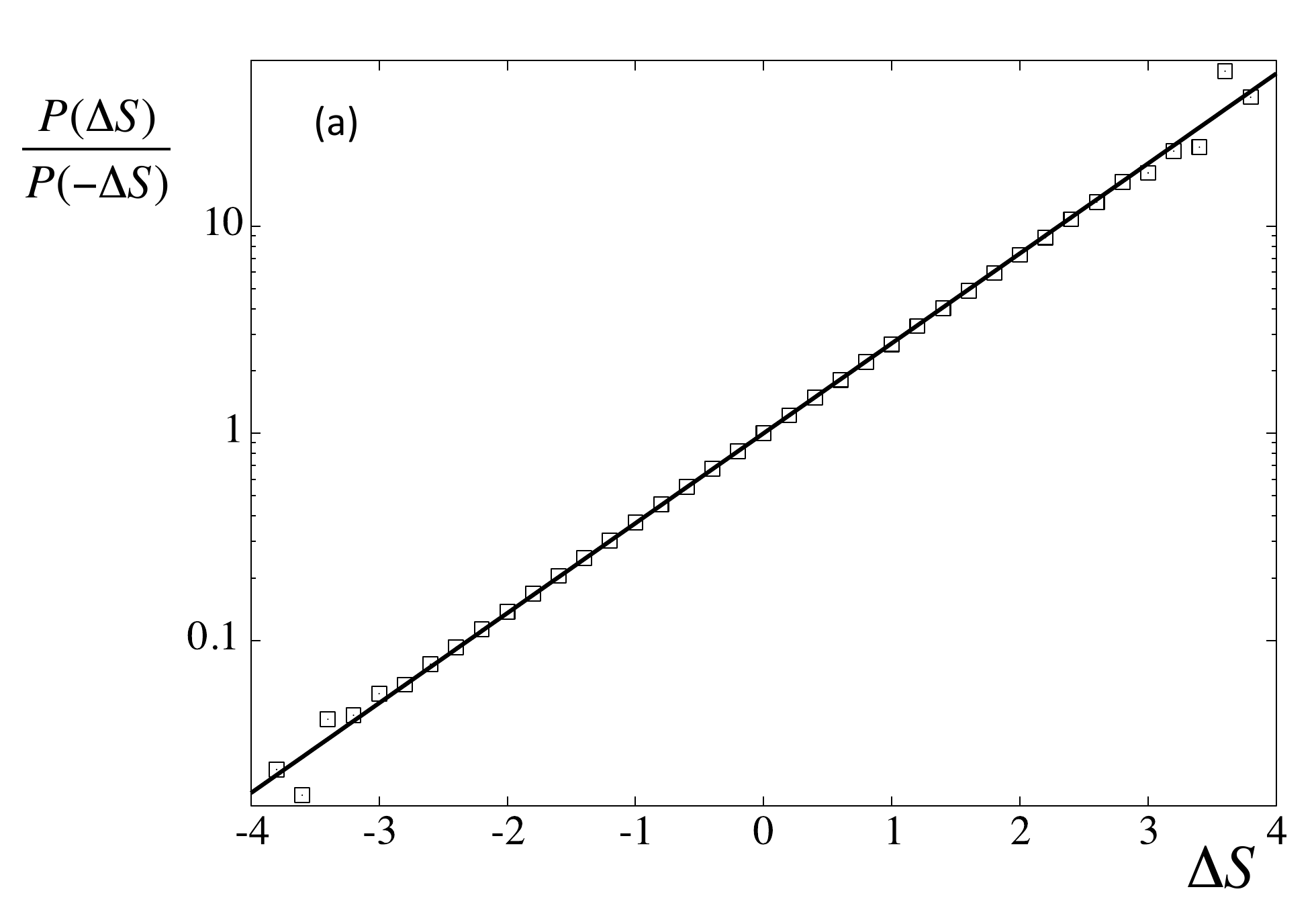}
\includegraphics[width=8cm,angle=0]{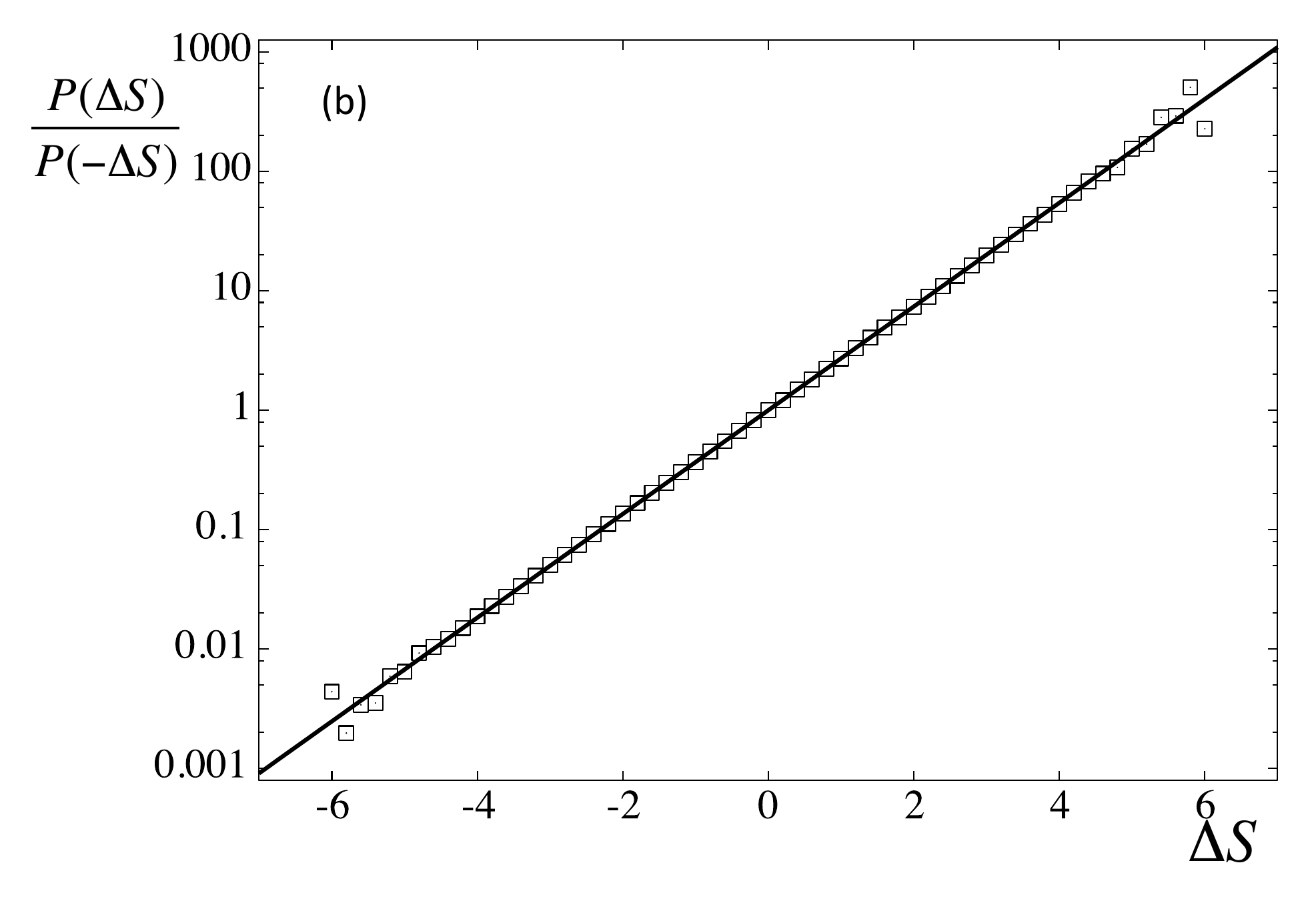}
\includegraphics[width=8cm,angle=0]{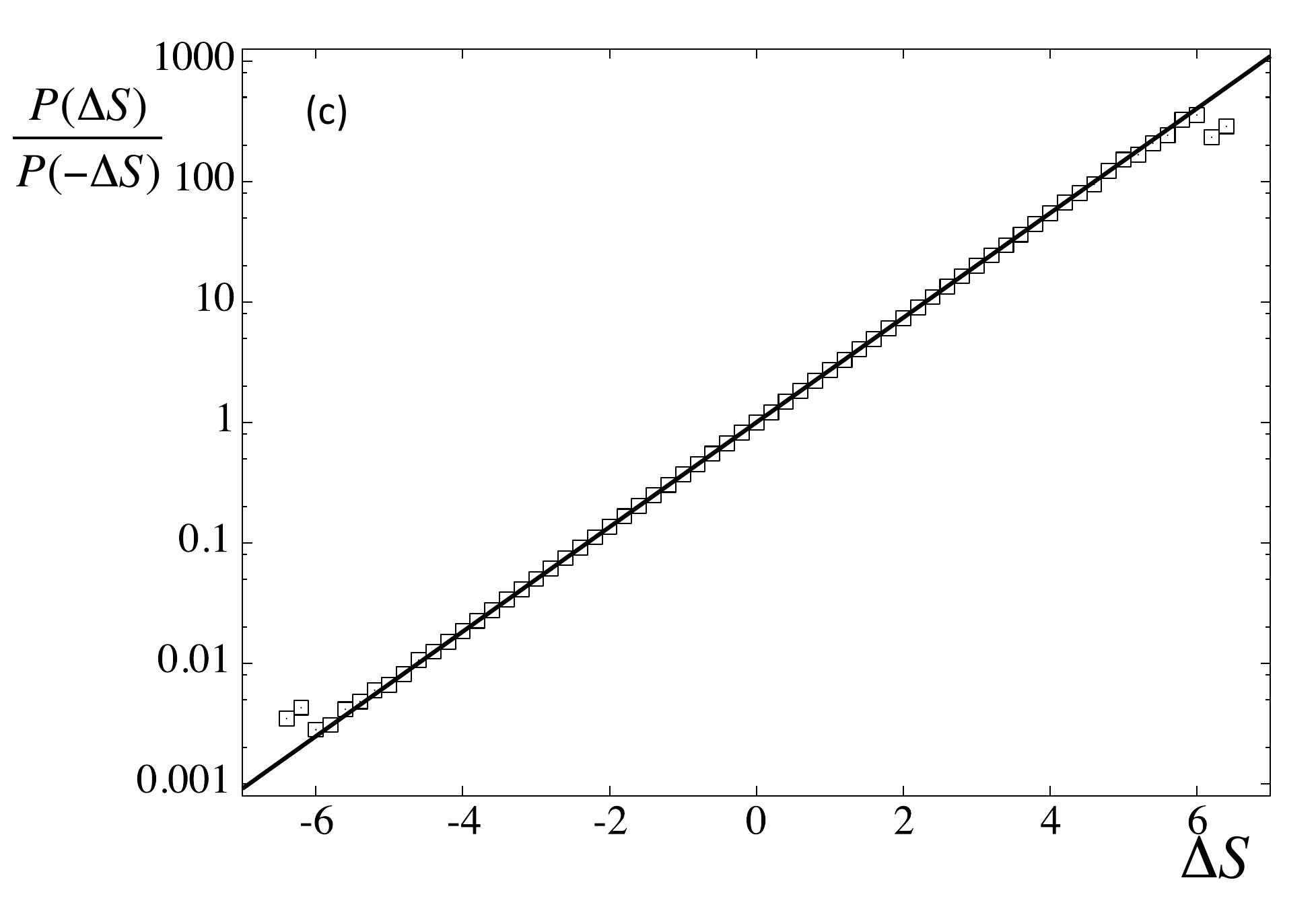}
\includegraphics[width=8cm,angle=0]{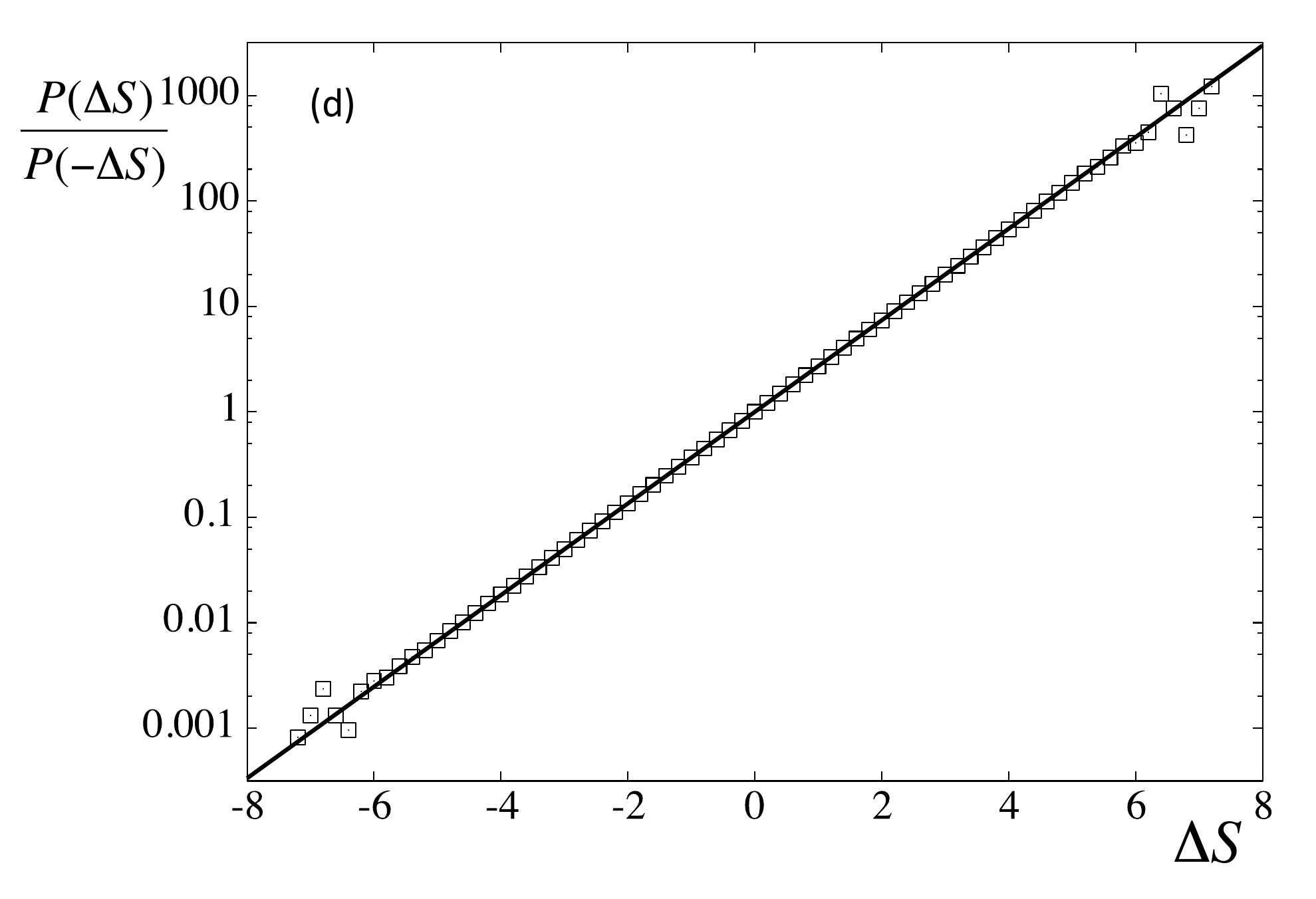}
\caption{\label{Fig:FT}Verification of the fluctuation theorem based on the data in Fig.~\ref{Fig:ps} (logarithmic scale on the vertical axis). The straight line corresponds to 
$\exp(\Delta S/k)$. Same parameter values as in Fig.~\ref{Fig:ps}.}
\end{figure}

With these preliminaries, the numerical verification of the fluctuation theorem Eq.~(\ref{FT}) proceeds according to following steps. Starting from an initial equilibrated configuration at time $t=0$, $\{s(0)\}$, we simulate numerically the stochastic process up to a time $t$ MCS (Monte Carlo steps), ending in a configuration $\{s(t)\}$. During this run, we monitor the amount $Q_2(t)$ of heat taken {\it from} $\mathbf{B_2}$ and an amount $-Q_1(t)$ of heat given {\it to} $\mathbf{B_1}$. The reservoir entropy production of this single realization of the stochastic process is given by:
\begin{equation}
\label{eq:SReservoir}
\Delta S_\textrm{Bath}(t)=\frac{-Q_2(t)}{T_2}+\frac{Q_1(t)}{T_1}.
\end{equation}
As the stochastic entropy is again a state function (but now of the stochastic state of the system), the change in system entropy for the run under consideration is  the final value minus the initial value: $\Delta S_\textrm{System}(t)/k=-\ln P_\textrm{st}(\{s(t)\})+\ln P_\textrm{st}(\{s(0)\})$.
The total entropy production then follows as the sum of the reservoir and system contribution: $\Delta S(t)=\Delta S_\textrm{Bath}(t)+\Delta S_\textrm{System}(t)$. 
By generating a large number of runs and recording the corresponding values of $\Delta S(t)$, one can construct a histogram for $P(\Delta S,t)$ . The results are in excellent agreement with the fluctuation theorem as shown in the Figs.~\ref{Fig:ps} and \ref{Fig:FT} \ for $M=10$ with $t=10$, $T_1=1$ and  $T_2=2,5,10,\infty$. Similar results are obtained for smaller and larger system sizes. As the system size increases, one notes that the system contribution to the entropy being bounded, becomes less important, and the fluctuation theorem converges to its time-asymptotic formulation, involving only the reservoir contribution. As an independent check of the simulations, we have also verified that, by
averaging over many realizations, we reproduce the aforementioned average heat flux and corresponding reservoir entropy production $\langle \Delta S_\textrm{Bath}(t)\rangle =tJ_Q\displaystyle \left(\frac{1}{T_1}-\frac{1}{ T_2}\right)>0$.

Note finally a peculiar property of the probability distribution for the stochastic
entropy: while obeying the fluctuation theorem, $P(\Delta S)$ does have an unexpected shape with several bizarre peaks, cf. Figs. \ref{Fig:ps}, a feature that disappears in the limit of a large system size. A similar phenomenon has been observed in other stochastic models with discrete step-like dynamics, notably in a single level quantum dot  \cite{Esposito:2007}. Unfortunately, the explicit analytic expression of the stochastic entropy cannot be obtained even for the simplest case of two spins discussed below, and the precise nature of this feature remains to be elucidated.

\section{The large deviation function for the single-particle case}
\label{sec:large}
After having presented mostly numerical data about the stochastic particle flux, heat flux and entropy production for a finite system we finally present the exact analytic result for the case of only two spins, corresponding, in the particle interpretation, to a single site allowing at most one particle. This study is complimentary to the analysis of other two-state systems \cite{Lacoste:2008,Esposito:2009a,Willaert:2014,Cornu:2013}, to the study of particle transported in models without particle interaction \cite{Harbola2014,VdB2012}, and to exact asymptotic results in the limit of very large systems sizes \cite{derrida,Fogedby:2012}. Our exact results allows to compare in detail the short and intermediate time behavior with the asymptotic large time behavior embodied in the large deviation function. 

We focus on the the large deviation function, describing the asymptotic time regime. In this way, one can evaluate the finite time probability distribution for the stochastic entropy production $P(\Delta S)$. Of particular interest to us is how the fluctuation theorem goes over into its asymptotic form in which both the system contribution and the effect of the initial preparation disappear. We expect that this will be the case after a few time steps as the system entropy is limited to $kT\ln2$.
While the analytic result for $P(\Delta S)$ at finite times are still quite complicated, the large deviation function is relatively simple. It can be obtained by the following short-cut. For long times, we will have that the stochastic entropy production is given by $\Delta S=(\mu_2/T_2-\mu_1/T_1)Jt$, where $J=N/t$ is the stochastic particle current, defined now from reservoir $\mathbf{B_1}$ into the site. Note that we are neglecting here the entropy of the system, and the fact that the current into the system can differ by $\pm1$ from the current between the reservoirs. Hence it is sufficient to evaluate the large deviation of the current $J$.

Our starting point considers the probability distribution for both the state of the system and the net number of particles $N$ that have been injected from the $\mathbf{B_1}$ reservoir during a time $t$, namely, $P_0(N;t)\equiv P(N;\tau_1=0;t)$ and $P_1(N;t)\equiv P(N;\tau_1=1;t)$, with the subscript $0$ and $1$ referring to whether there no particle or a single particle in the site. They satisfy the master equation:
\begin{eqnarray}
\label{p0n1n2}
\frac{\partial P_0(N;t)}{\partial t}&=&\gamma P_1(N+1;t)+\beta P_1(N;t)-(\alpha+\delta)P_0(N;t),\\
\label{p1n1n2}
\frac{\partial P_1(N;t)}{\partial t}&=&\alpha P_0(N-1;t)+\delta P_0(N;t)-(\beta+\gamma)P_1(N;t).
\end{eqnarray}
The probability $P(N;t)$ of interest, i.e., for having a cumulative number of particles $N$, or a corresponding flux of $J=N/t$ from reservoir $\mathbf{B_1}$ into the system after a time $t$, is obtained by summing out the state of the system $P(N;t)=P_0(N;t)+P_1(N;t)$. 

\begin{figure}
\includegraphics[width=14cm,angle=0]{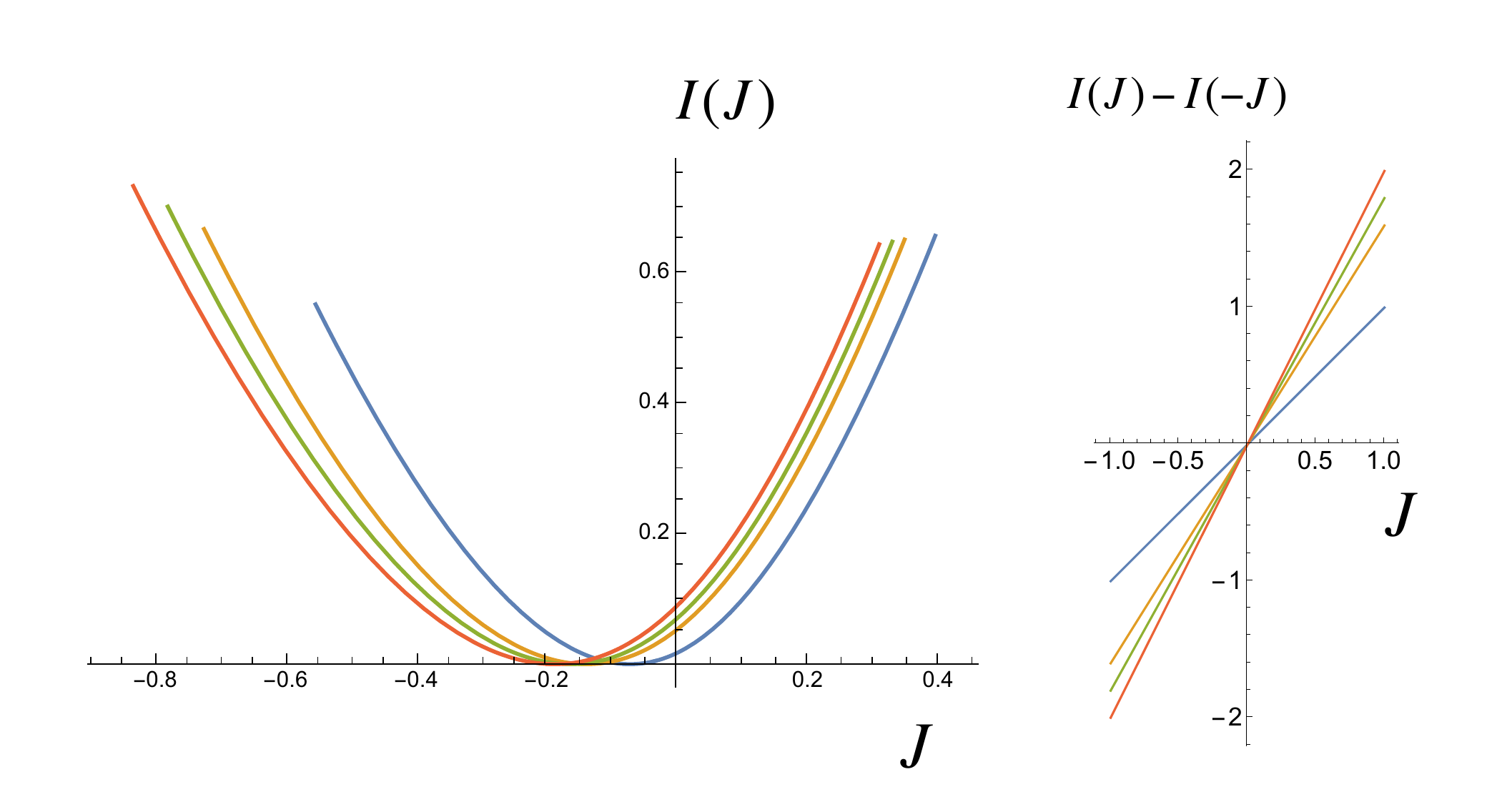}
\caption{\label{Fig:large}Left panel:  large deviation function obtained by parametric elimination of $\xi_m$ from Eqs.~(\ref{eq:J1}) and (\ref{eq:J2}). Here, $u(\xi)$ is given by 
Eq.~(\ref{eq:uxi}) and the exchange parameters are chosen according to Eqs.~(\ref{eq:equivalence}) and (\ref{psfp}). We set $\lambda=1$, and $\epsilon=2,\,k=1$ as before. The $4$ different curves, from right to left, correspond to  $T_1=1$ and $T_2=2, 5, 10, \infty$, respectively. Right panel:  $I(J)-I(-J)$  versus $J$, for the same temperature values. These are, in agreement with the fluctuation theorem,  linear functions of $J$ with slope  equal to $\dfrac{\epsilon}{k}\left(\dfrac{1}{T_1}-\dfrac{1}{T_2}\right)$, cf. Eq.~(\ref{FTLD}).}
\end{figure}

Eqs.(\ref{p0n1n2}-\ref{p1n1n2}) can be solved by introducing the generating functions
\begin{eqnarray}
G_0(\xi;t)&=&\sum_{N}e^{\xi N}P_0(N;t),\\
G_1(\xi;t)&=&\sum_{N}e^{\xi N}P_1(N;t).
\end{eqnarray}
They verify:
\begin{eqnarray}
\frac{\partial G_0(\xi,t)}{\partial t}&=& -(\alpha+\delta)G_0(\xi,t)+\left(\beta+\gamma e^{-\xi}\right)G_1(\xi,t),\\
\frac{\partial G_1(\xi,t)}{\partial t}&=& -(\gamma+\beta)G_1(\xi,t)+(\delta +\alpha e^{\xi}) G_0(\xi,t).
\end{eqnarray}
Note that $\xi$ is just a parameter in these equations. 
Therefore, we have a system of two ordinary (not partial) differential equations. 
After some algebra, the solution satisfying the initial condition $G_0(\xi,0)=1,\,G_1(\xi,0)=0$ corresponding to starting with no particle in the system at $t=0$, can be written as:
\begin{eqnarray}
G_0(\xi,t)&=&e^{-\lambda t}\left[\cosh\left(\lambda t \sqrt {u(\xi)}\right)+(\gamma-\delta)\frac{\sinh\left(\lambda t\sqrt{u(\xi)}\right)}{\sqrt{u(\xi)}}\right],\\
G_1(\xi,t)&=&e^{-\lambda t}\frac{\sinh\left(\lambda t\sqrt{u(\xi)}\right)}{\sqrt{u(\xi)}}(\delta+\alpha e^{\xi}),\\
u(\xi)&\equiv&(\delta+\beta e^{\xi})\left(\alpha+\gamma e^{-\xi}\right),\label{eq:uxi}
\end{eqnarray}
and we have used $\alpha+\gamma=\beta+\delta=\lambda$. To extract the large deviation function of the current, we first derive from the above exact expression the asymptotic behavior of the cumulant generating function $G=G_0+G_1$:
\begin{eqnarray}
G(\xi)\sim e^{-\lambda t \left(1-\sqrt{u(\xi)}\right)}.
\end {eqnarray}
The large deviation function $I(J)$ quantifies the exponentially small probability for observing a current $J=N/t$ in the large $t$-limit :
\begin{eqnarray}
P(N,t)\sim e^{-tI(J)}.
\end{eqnarray}
It is related by Legendre transform to the asymptotic behavior of the cumulant generating function (since the latter is continuous differentiable \cite{Touchette:2009}):
\begin{eqnarray}\label{eq:IJ}
I(J)=\min_{\xi}\left\{J\xi+\lambda\left(1-\sqrt{u(\xi)}\right)\right\}.
\end{eqnarray}
The minimum is reached for $\xi_m$ obeying:
\begin{eqnarray}\label{eq:J1}
J&=&\frac{\lambda}{2}\frac{u'(\xi_m)}{\sqrt{u(\xi_m)}},
\end{eqnarray}
hence
\begin{eqnarray}\label{eq:J2}
I(J)&=&J\xi_m+\lambda\left(1-\sqrt{u(\xi_m)}\right).
\end{eqnarray}
The large deviation function $I(J)$ is then readily obtained by parametric elimination of $\xi_m$ from these two equations, see Fig.~\ref{Fig:large}.

Turning to the fluctuation theorem, we note that a particle current $J$ produces a entropy production rate $J_S$ given by:
\begin{eqnarray}
J_S=X J\\
X=\frac{\epsilon-\mu_1}{T_1}-\frac{\epsilon-\mu_2}{T_2}
\end{eqnarray}
As stated before, we neglect here the fact that a particle may have entered the system from one reservoir without moving into the other reservoir, as well as the bounded contribution of the entropy
production in the system. Hence the large deviation properties of the entropy production are identical to those of the current, apart from the rescaling by the prefactor $X$.
This factor can be interpreted as the effective thermodynamic force. The fact that there is a single thermodynamic force while there are two gradients (in chemical potential and 
in temperature) is a result of the strong coupling of the particle and energy flux (hence $J_S=X_Q J_Q+XJ$ reduces to the above expression).

To make the connection with the fluctuation theorem for the entropy production of the reservoirs in the large $t$ limit, we note that $u(\xi)$, and hence the cumulant generating function $\phi(\xi)=-t^{-1}\ln G(\xi)$, is invariant under the transformation $\xi \rightarrow \xi_0-\xi$ with $\xi_0=\ln\left[\dfrac{\delta\gamma}{\alpha\beta}\right]=\ln\left[\dfrac{p_2(1-p_1)}{p_1(1-p_2)}\right]=x_1-x_2=X/k$, which is precisely the thermodynamic force $X$ divided by Boltzmann's constant. Since the large deviation function $I(J)$ is related to the cumulant generating function by Legendre transform $I(J)=\textrm{ext}_{\xi}\left\{\phi(\xi)+\xi J\right\}$, one concludes that:
\begin{equation}\label{FTLD}
I(J)=I(-J)+XJ/k,
\end{equation}
which is the expression of the fluctuation theorem in terms of the large deviation function of the current.

\section{Discussion}
\label{sec:discussion}

Stochastic thermodynamics provides the generalization of thermodynamics to the description of small nonequilibrium systems. In this paper we have studied in the novel context of stochastic thermodynamics, the one dimensional Ising model and the simple symmetric exclusion process. These models  are  among the best studied models in (non)equilibrium statistical mechanics and are particularly well suited for investigating the system contribution to the stochastic entropy, as it is one of the very few cases for which the nonequilibrium steady state  probability is known exactly.  Our results provide yet another illustration of the powerful formalism of stochastic thermodynamics, with an application to a spatially extended system obeying micro-canonical dynamics. We verify two specific predictions, namely the universality of efficiency at maximum power for thermal machines in the simple symmetric exclusion process  and the fluctuation theorem for the finite Ising chain in contact with two thermal reservoirs.


Acknowledgements: 
R.T. acknowledges support from by FEDER (EU) and MINECO (Spain) under Grant ESOTECOS FIS2015-63628-C2-R, D. E. from FONDECYT Project No. 1140128, C. VdB. from COST Action MP1209 STSM 34396 and K.L. from the U.S. Office of Naval Research (ONR) under Grant No. N00014-13-1-0205.

\section*{Appendix I: Numerical simulation of the Master equation for the Ising chain and calculation of the entropy change}

In our numerical simulations we use a discrete-time Monte Carlo update scheme\cite{Toral-Colet:2014}. We first randomly select a node $i=1,\dots,M$. Then: \\
\noindent-If the node is $i=1$ or $i=M$ we replace the spin variable $s_1$ or $s_M$ by a new value $\pm 1$ chosen with the heat-bath probabilities ($k$ is Boltzmann's constant)
\begin{eqnarray}
\label{ps1}
\textrm{prob}(s_1=\pm 1 )=\frac{1}{1+e^{\mp s_2\epsilon/kT_1}}\textrm{~~~~~~~~~~prob}(s_M=\pm 1 )=\frac{1}{1+e^{\mp s_{M-1}\epsilon/kT_2}}.
\end{eqnarray}
Note that, correspondingly, the link variables $\ell_1=-s_1s_2$ and $\ell_{M-1}=-s_{M-1}s_M$ can take two values with probabilities:
\begin{eqnarray}
\label{psa1}
\textrm{prob}(\ell_1=\pm 1)=\frac{1}{1+e^{\pm \epsilon/kT_1}},\,\textrm{prob}(\ell_{M-1}=\pm 1)=\frac{1}{1+e^{\pm \epsilon/kT_2}}.
\end{eqnarray}
To simplify the notation, we use the shorthand:
\begin{equation}
\label{psf}
p_{1,2}=1-q_{1,2}=\frac{1}{1+e^{\epsilon/kT_{1,2}}}.
\end{equation}

\noindent-If the chosen node satisfies $1<i<M$, we use a microcanonical update: since the contribution of spin $i$ to the total energy is $-s_i(s_{i+1}+s_{i-1})$, the flip $s_i\to-s_i$ is accepted if and only if $s_{i-1}+s_{i+1}=0$. 

This elementary update is repeated $t$ Monte Carlo steps (defined as $M$ single spin update trials). We denote the value of the spin $s_i$ after the single spin update trial number $n$ by $s_i(n)$, where $n=1,\dots,Mt$. At a selected time $t$ we compute the heat fluxes and the change of entropy of the reservoir using Eq.~(\ref{eq:SReservoir}). For this, we first compute the heat $Q_2(t)$ taken from $\mathbf{B_2}$ during the time interval $(0,t)$ (starting to count after the equilibration updates). This is defined as the following sum over spin updates:
\begin{equation}
Q_2(t)=\frac{\epsilon}{2}\times\sum_{\textrm{updates $n$ where $s_M$ has been selected}}\Delta \ell_M(n)
\end{equation}
with $\Delta \ell_M(n)=(s_M(n-1)-s_M(n))s_{M-1}(n-1)$, the energy change due only to updating variable $s_M$. Similarly, the heat $Q_1(t)$ taken from reservoir $\mathbf{B_1}$ is defined as:
\begin{equation}
Q_1(t)=-\frac{\epsilon}{2}\times\sum_{\textrm{updates $n$ where $s_1$ has been selected}}\Delta \ell_1(n)
\end{equation}
with $\Delta \ell_1(n)=(s_1(n-1)-s_1(n))s_{2}(n-1)$, the energy change due only to updating variable $s_1$. Note that the definition is such that $Q_{1,2}(t)<0\,(\textrm{resp. }>0)$ when energy is given to (resp. taken from) the respective reservoirs $\mathbf{B_1}$ and $\mathbf{B_2}$. For the average over realizations we expect that $-\langle Q_1(t)\rangle=\langle Q_2(t) \rangle\equiv \langle Q(t) \rangle>0$ ($T_2>T_1$). 

\section*{Appendix II: Calculation of the flux}
We first compute the probabilities of the possible values of $\Delta \ell_1$ in a single spin update. The link energy $\ell_1$ is allowed to change only when $s_1$ is selected, a process that occurs with probability $1/M$. There are four possibilities:\\
1) If $s_1=s_2=+1$, the change is $\Delta \ell_1=+2$ only if $s_1$ changes to $s_1=-1$ (an event with probability $p_1$), otherwise the change is $0$.\\
2) Similarly, if $s_1=s_2=-1$, the change is $\Delta \ell_1=+2$ only if $s_1$ changes to $s_1=1$, an event with probability $p_1$.\\
3) If $s_1=+1,\,s_2=-1$, the change is $\Delta \ell_1=-2$ only if $s_1$ changes to $s_1=-1$, an event with probability $q_1$.\\
4) Finally, the case $s_1=-1,\,s_2=+1$, leads to a change $\Delta \ell_1=-2$ with probability $q_1$. 

We add all contributions and write them in terms of the reduced stationary probability distribution $P_\textrm{st}(s_1,s_2)=\sum_{s_3,s_4,\dots,s_M} P_\textrm{st}(s_1,s_2,\dots,s_M)$. 
\begin{eqnarray}
\textrm{Prob}(\Delta \ell_1=+2)&=&\frac{1}{M}P_\textrm{st}(1,1)p_1+\frac{1}{M}P_\textrm{st}(-1,-1)p_1,\\
\textrm{Prob}(\Delta \ell_1=-2)&=&\frac{1}{M}P_\textrm{st}(1,-1)q_1+\frac{1}{M}P_\textrm{st}(-1,1)q_1.
\end{eqnarray}
Due to the symmetry of the problem, we have $P_\textrm{st}(1,1)=P_\textrm{st}(-1,-1)$ and $P_\textrm{st}(-1,1)=P_\textrm{st}(1,-1)$. Using the normalization condition $\sum_{s_1=\pm1,s_2\pm 1}P_\textrm{st}(s_1,s_2)=1$, it turns out that the average energy taken from $\mathbf{B_1}$ during a time interval $t=1/M$ (a single spin update) is
\begin{equation}
\label{q1av}
\langle Q_1(t=1/M)\rangle=-\frac{\epsilon}{2}\times\frac{2}{M}\left[2P_\textrm{st}(-1,1)-p_1\right]=\frac{\epsilon}{M}\left[P_\textrm{st}(\ell_1=+1)-p_1\right].
\end{equation}
The exact solution \cite{Spohn:1983} shows that the probabilities of the different energies of the link $s_is_{i+i}$ follow a linear dependence on the distance to the reservoirs:
\begin{eqnarray}
\label{pstepsi}
P_\textrm{st}(\ell_i=+1)&=&\frac{(M-i)p_1+ip_2}{M}.
\end{eqnarray}
Taking $i=1$ and substituting in Eq.(\ref{q1av}), we obtain $\langle Q_1(t=1/M)\rangle=-\dfrac{\epsilon(p_2-p_1)}{M^2}$, and a current $J_Q=-\langle Q_1(t=1/M)\rangle/(1/M)=\dfrac{\epsilon(p_2-p_1)}{M}$. Using $J_Q=\epsilon J$ we conclude that the particle flux is: 
\begin{equation}
J=\frac{p_2-p_1}{M}.
\end{equation}
Comparing with the equivalent result of the symmetric exclusion process, we conclude that the time scale factor must be set to $\lambda=1$ to reproduce the discrete-time simulation results.

If the temperature difference between the two ends of the chain is small $\Delta T=T_2-T_1\ll 1$, then it is possible to expand the current:
\begin{equation}
J_Q=\frac{\epsilon}{M}(p(T_1+\Delta T)-p(T_1))=\frac{\epsilon}{k}\left(2T_1\cosh(\epsilon/2kT_1)\right)^{-2}\dfrac{\Delta T}{M}+O(\Delta T)^2,
\end{equation}
where $p(T)=\dfrac{1}{1+e^{\epsilon/kT}}$. This is simply Fourier's law in its simplest version that the current is proportional to the temperature gradient. Far away from this linear regime, the verification of Fourier's law requires the introduction of a local temperature $T(x)$. This can be achieved (in the steady state) by setting the probability of link $\ell_i$ to have energy $ \epsilon/2$ as $\textrm{prob}(\ell_i=1)=p(T_i)$,  which combined with Eq.(\ref{pstepsi}) leads to 
\begin{equation}
p(T_i)=\frac{(M-i)p(T_1)+ip(T_2)}{M},
\end{equation}
or in terms of continuous variables $x=i\Delta x, \, L=M\Delta x$
\begin{equation}
p(T(x))=\frac{(L-x)p(T_1)+xp(T_2)}{L},
\end{equation}
which defines the temperature profile as
\begin{equation}
T(x)=\dfrac{\epsilon/k}{\log\left(\dfrac{L}{(L-x)p(T_1)+xp(T_2)}-1\right)}.
\end{equation}

Now it is possible to satisfy Fourier law (at least in the steady state) introducing a suitable heat conductivity $\kappa(x)$ such that $J_Q=\kappa(x)\dfrac{dT(x)}{dx}$. Using $J_Q=\epsilon J$ and the afore-defined $T(x)$ one finds after a simple algebra:
\begin{equation}
\kappa(x)=\epsilon\dfrac{dp(T)}{dT}=\frac{\epsilon}{k}\left[2T(x)\cosh(\epsilon/2kT(x))\right]^{-2},
\end{equation}
independent of system size $L$.

\section*{Appendix III: Calculation of the expansion of the efficiency at maximum power}
We start from
\begin{equation}
{\cal P}=(x_2-(1-\eta_C)x_1)[f(x_2)-f(x_1)]
\end{equation}
and will later specify the appropriate form of the function $f(x)$. To find the values  $(x_1^*,x_2^*)$ that maximize $\cal P$, we need to solve the equations:
\begin{eqnarray}
\left.\frac{\partial {\cal P}}{\partial x_1}\right|_{(x_1^*,x_2^*)}&=&(-1+\eta_C)f(x_2^*)-f(x_1^*)-(x_2^*-x_1^*(1-\eta_C)f'(x_1^*)=0,\\
\left.\frac{\partial {\cal P}}{\partial x_2}\right|_{(x_1^*,x_2^*)}&=&f(x_2^*)-f(x_1^*)+(x_2^*-x_1^*(1-\eta_C)f'(x_2^*)=0.
\end{eqnarray}
Inserting the expansions (\ref{x1a0}-\ref{x2b0}) we obtain at order $\eta_C^0$ that $a_0=b_0$. At order $\eta_C^1$ one finds:
\begin{equation}
b_1=a_1+a_0/2.
\end{equation}
It is only when going to order $\eta_C^2$ that $a_0$ is found as the solution of the equation:
\begin{equation}
2 f'(a_0)+a_0f''(a_0)=0,
\end{equation}
with in addition:
\begin{equation}
b_2=a_2+\frac{a_1}{2}+\frac{3a_0}{8}.
\end{equation}
At this order $a_1$ and $a_2$ are still not determined. Note however that  by expanding
\begin{equation}
\eta^*=1-(1-\eta_C)\frac{x_1^*}{x_2^*}=1-(1-\eta_C)\frac{a_0+(a1+a_0/2)\eta_C+(a_2+\frac{a_1}{2}+\frac{3a_0}{8})\eta_C^2}{a_0+a_1\eta_C+a_2\eta_C^2}=\frac{\eta_C}{2}+\frac{\eta_C^2}{8}+O(\eta^3),
\end{equation}
we reproduce the known universal coefficients $1/2$ and $1/8$, irrespective of the values of  $a_0,a_1,a_2$ and of the function $f(x)$.
At order $\eta_C^3$  we find 
\begin{eqnarray}
a_1&=&-\frac{a_0}{4},\\
b_3&=&a_3+\frac{a_2}{2}+\frac{3a_0}{16}-\frac{a_0^2 f'''[a_0]}{96 f''[a_0]},
\end{eqnarray}
determining the value of  $a_1$ and hence $b_1$. At order $\eta_C^4$ we find $a_2$ as a function of $a_0$ (and hence we can determine $b_2$), and $b_4$ as a function of $a_0,a_3, a_4$. It is only at order $\eta_C^5$ that we find the explicit values of $a_3$ and $b_3$. 

In summary, to find the coefficients of the expansion of $x_2^*,x_1^*$ to order $\eta_C^k$ we need to go to order $\eta_C^{k+2}$, but given the relations between coefficients it turns out that the expansion for $\eta^*$ in (\ref{etan}) is correct to order $\eta_C^{k+2}$ . Specifically, for $f(x)=1/(1+e^x)$ we find
\begin{eqnarray}
x_1^*&=&2.39936 + 0.599839 \eta_C + 0.399893 \eta_C^2 + 0.294431 \eta_C^3 + 0.230513 \eta_C^4+\dots\\
x_2^*&=&2.39936 - 0.599839 \eta_C - 0.199946 \eta_C^2 - 0.0944843 \eta_C^3 - 0.0529399 \eta_C^4+\dots
\end{eqnarray}
from where we obtain Eq.(\ref{etaexp}).

\section*{Appendix IV: Calculation and properties of the stationary distribution}
Given the isomorphism between the Ising and the particle versions of the model, it is possible to use the exact result for the stationary distribution as found in \cite{Derrida:2002}:
\begin{equation}
\label{exactpst}
P_\textrm{st}(\tau_1,\dots,\tau_L)=\frac{\langle W|\prod_{i=1}^L(\tau_iD+(1-\tau_i)E)|V\rangle}{\langle W|(D+E)^L|V\rangle}
\end{equation}
where the operators $E,\,D$ and the vectors $|V\rangle,\,|W\rangle$ are defined by:
\begin{eqnarray}
[D,E]\equiv DE-ED&=&D+E,\\
(\beta D-\delta E)|V\rangle&=&|V\rangle,\\
\langle W|(\alpha E-\gamma D)&=&\langle W|.
\end{eqnarray}
The idea is simple, given a configuration $\{s\}=(s_1,\dots,s_{M})$, translate into $\{\tau\}=(\tau_1,\dots,\tau_L)$ with $L=M-1$ and then apply the above formula and $P_\textrm{st}(\{s\})=\frac12P_\textrm{st}(\{\tau\})$, as one configuration $(\tau_1,\dots,\tau_{L})$ is equivalent to two configurations $(s_1,\dots,s_M)$ which differ only on a global sign. For instance, the configuration $\{s\}=(-1,1,-1,1,1)$ corresponds to $\{\tau\}=(1,1,1,0)$ whose probability is:
\begin{equation}
P_\textrm{st}(1,1,1,0)=\frac{\langle W|DDDE|V\rangle}{\langle W|(D+E)^4|V\rangle}.
\end{equation}
To compute this, we use the following algebra: define 
\begin{eqnarray}
X&=&\beta D-\delta E \Rightarrow X|V\rangle=|V\rangle,\\
Y&=&\alpha E-\gamma D \Rightarrow \langle W|Y=\langle W|.
\end{eqnarray}
The commutator of $X$ and $Y$ and the inverse relations are:
\begin{eqnarray}
\label{comxy}
[X , Y] &=&X+Y,\\
E&=&\frac{\gamma X+\beta Y}{\alpha\beta-\gamma\delta},\\
D&=&\frac{\alpha X+\delta Y}{\alpha\beta-\gamma\delta},
\end{eqnarray}
where we have used $\alpha+\gamma=\beta+\delta=\lambda=1$.

In practice one defines rescaled operators 
\begin{eqnarray}
\label{resce}
\hat{E}&=&\gamma X+\beta Y,\\
\label{rescd}
\hat{D}&=&\alpha X+\delta Y,
\end{eqnarray}
and uses the known value of the denominator of Eq.(\ref{exactpst}) to write:
\begin{equation}
\label{exactpst2}
P_\textrm{st}(\tau_1,\dots,\tau_L)=\frac{1}{(L+1)!}\frac{\langle W|\prod_{i=1}^L(\tau_i\hat{D}+(1-\tau_i)\hat{E})|V\rangle}{\langle W|V\rangle}
\end{equation}

The method to find $P_\textrm{st}(\{\tau\})$ is to write in this equation the operators $\hat{E},\,\hat{D}$ in terms of $X,\,Y$ using Eqs.(\ref{resce},\ref{rescd}), make repeated use of the commutation relation Eq.(\ref{comxy}) to get a sort of ``normal order'' in which all $X$'s are to the right of $Y$'s and then apply $X|V\rangle=|V\rangle,\,\langle W|Y=\langle W|$. 

The process is cumbersome to carry out in detail. It is possible to use non-commutative symbolic packages \footnote{We used the ones developed for Mathematica\cite{mathematica} by researchers at the UCSD Mathematics department and available at {\tt http://www.math.ucsd.edu/$\sim$ncalg/}.}  to do this algebra. However, we have not been able to obtain explicit expressions beyond system sizes $L=8$. For larger sizes, we turn to numerical methods to compute $P_\textrm{st}(\{s\})$. 

For larger values of $L\lesssim 25$ we have computed $P_\textrm{st}(\{s\})$ by solving numerically the stationary solution of the master equation (\ref{me}):
\begin{equation}
\label{pst}
P_\textrm{st}(\{s\})=\sum_{\{s'\}}\omega(\{s'\}\to\{s\})P_\textrm{st}(\{s'\}).
\end{equation}
This is equivalent to finding the eigenvector of eigenvalue $1$ of the transition matrix $\omega$. This matrix has, in principle, a dimension of $2^M\times 2^M$ (recall $M=L+1$). However, most of the entries are $0$ since the rules of the process only allow for transitions $\{s'\}\to\{s\}$ in which only one spin variable $s_i\to-s_i$ is changed. Therefore if $\{s\}=(s_1,\dots,s_M)$, the only configurations $\{s'\}$ for which $\omega(\{s'\}\to\{s\})$ is not equal to zero are, besides the configuration $\{s\}$ itself, the $M$ configurations $\{s\}_1\equiv(-s_1,s_2,s_3,\dots,s_M)$, $\{s\}_2\equiv(s_1,-s_2,s_3,\dots,s_M)$, $\dots$, $\{s\}_M\equiv(s_1,s_2,s_3,\dots,-s_M)$. The equation to solve is then:
\begin{equation}
P_\textrm{st}(\{s\})=\omega(\{s\}\to\{s\})P_\textrm{st}(\{s\})+\sum_{k=1}^M\omega(\{s\}_k\to\{s\})P_\textrm{st}(\{s\}_k).
\end{equation}
This equation we solve by iteration: take an initial guess in the right-hand side $P_\textrm{st}^{(0)}(\{s\})$ and iterate until there is convergence:
\begin{equation}
\label{pstn0}
P_\textrm{st}^{(n+1)}(\{s\})=\omega(\{s\}\to\{s\})P_\textrm{st}^{(n)}(\{s\})+\sum_{k=1}^M\omega(\{s\}_k\to\{s\})P_\textrm{st}^{(n)}(\{s\}_k).
\end{equation}
Note that this recursion relation strictly conserves the sum $\sum_{\{s\}} P_\textrm{st}^{(n)}(\{s\})=1$. We check that the numerical recursion conserves this normalization, something that we take as an indicator of its accuracy. We iterate until $\sum_{\{s\}} [P_\textrm{st}^{(n)}(\{s\})]^2$ does not vary significantly. We have found that convergence can be speeded up if, instead, we use the recursion relation:
\begin{equation}
\label{pstn}
P_\textrm{st}^{(n+1)}(\{s\})=\frac{\sum_{k=1}^M\omega(\{s\}_k\to\{s\})P_\textrm{st}^{(n)}(\{s\}_k)}{1-\omega(\{s\}\to\{s\})}.
\end{equation}
However, this does not conserve exactly the sum $\sum_{\{s\}} P_\textrm{st}^{(n)}(\{s\})=1$ and we add after (\ref{pstn}) the correction step:
\begin{equation}
\label{pstc}
P_\textrm{st}^{(n+1)}(\{s\})=\frac{P_\textrm{st}^{(n+1)}(\{s\})}{\sum_{\{s\}}P_\textrm{st}^{(n+1)}(\{s\})}.
\end{equation}

Both recursion relations converge to the same values and, for small system sizes, they also coincide to a high degree of accuracy (we have found agreement up to $16$ significant figures for $M \leq 8$) with the exact values obtained from the "Derrida solution" \cite{Derrida:2002}. This recursion method allows to compute the stationary distribution up to $M\approx 25$. 

For larger values of $M\gtrsim 25$ this method takes too long to converge. In this case, we have generated numerically the stationary distribution by running the Monte Carlo simulation for a sufficiently long time. The caveat of this method is that some configurations (specially those with high and small values of the energy) have a small probability and do not appear in a typical simulation run. Hence, the data for those large values of $M$ is not accurate at both ends of the energy scale.

\begin{figure}
\includegraphics[width=8cm,angle=0]{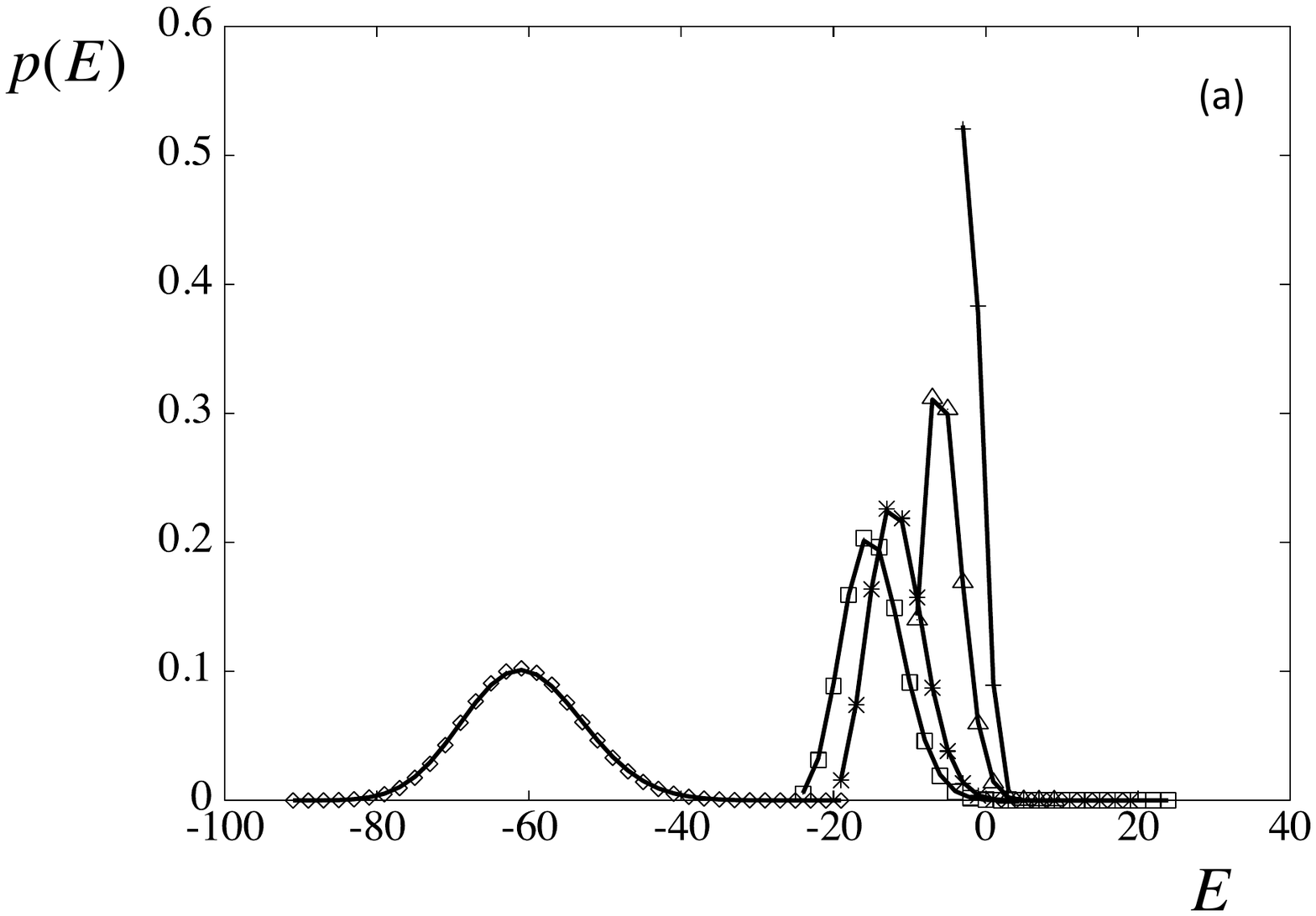}
\includegraphics[width=8cm,angle=0]{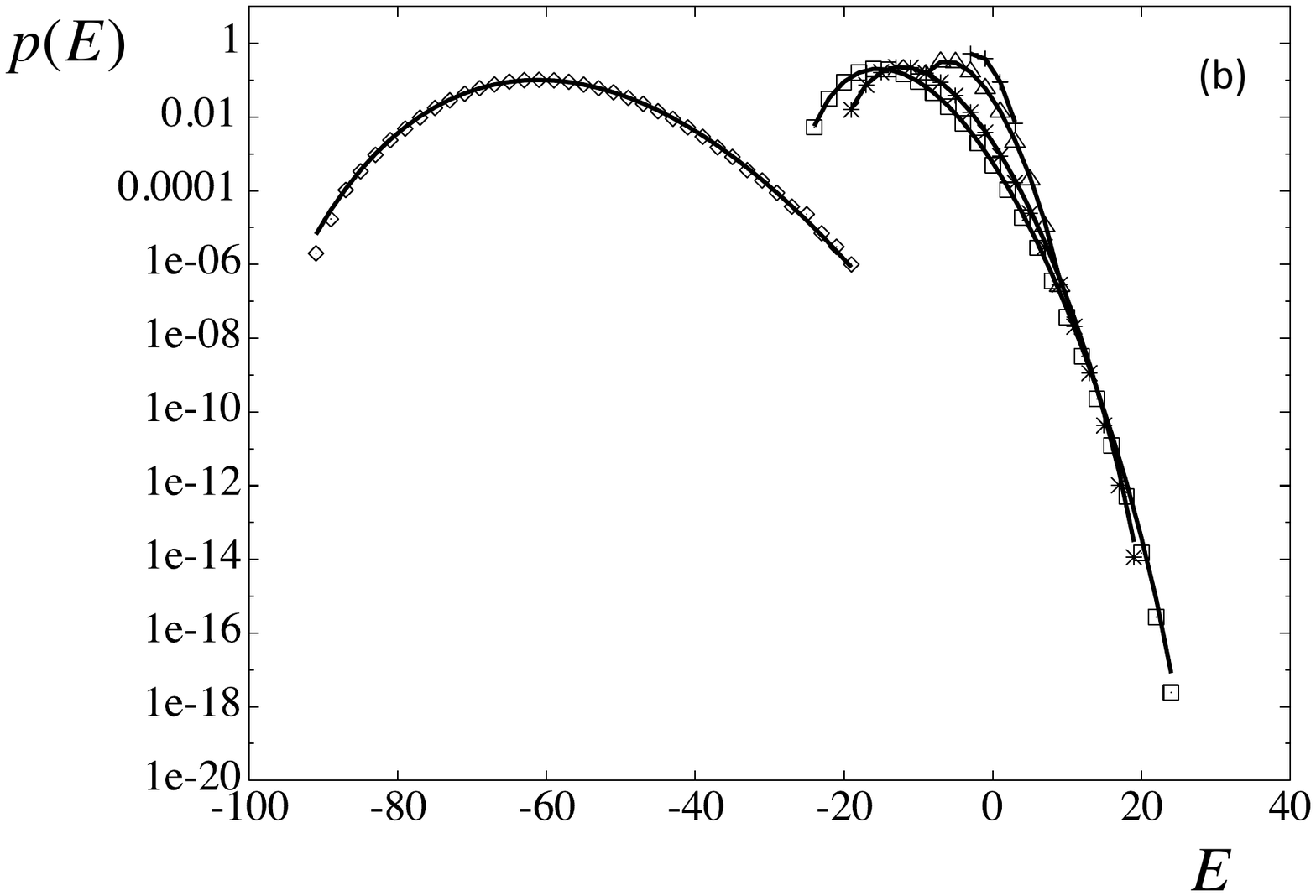}
\includegraphics[width=8cm,angle=0]{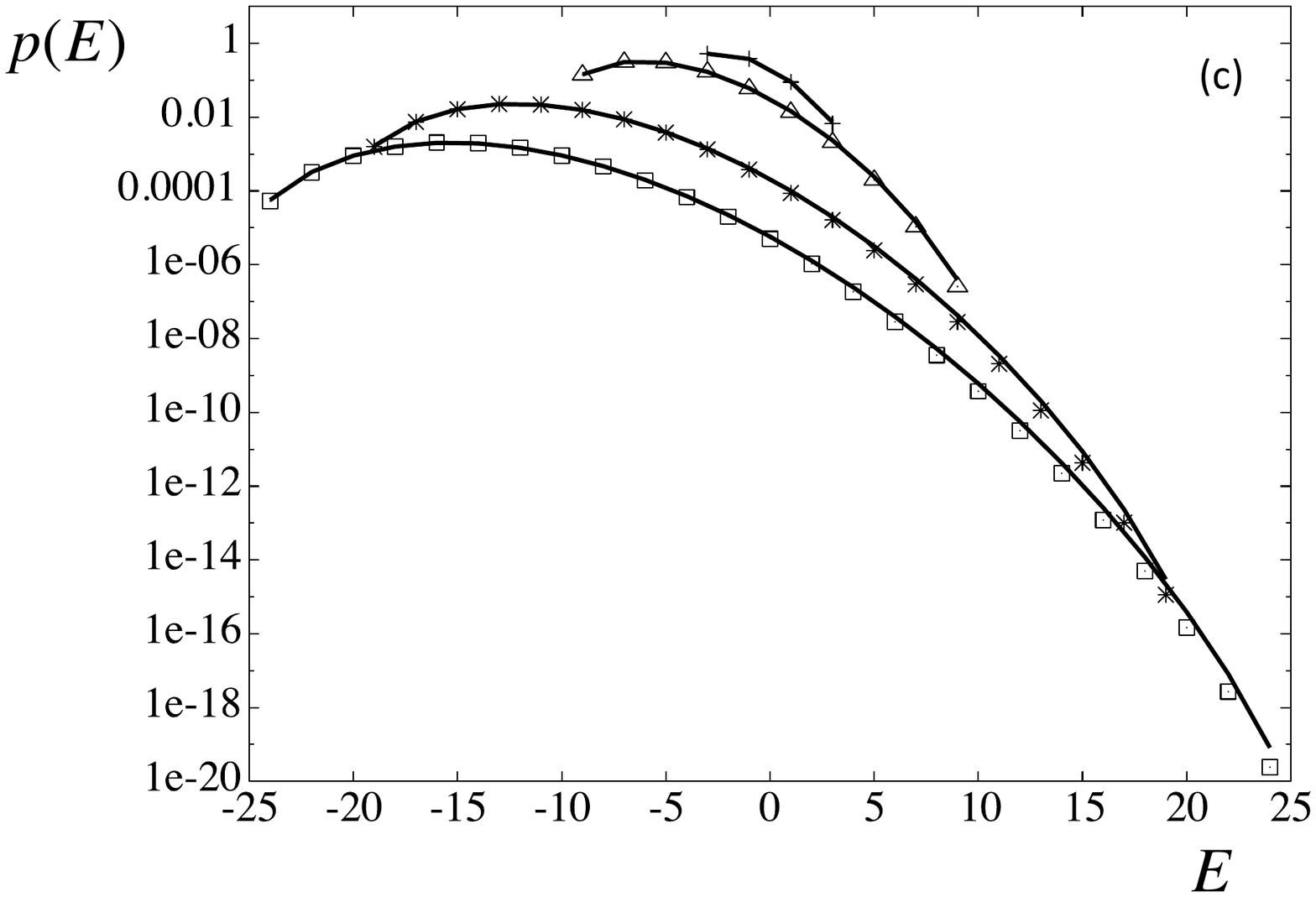}
\caption{\label{fig:peff}Fit of the effective form Eq.~(\ref{poeff}) to $p(E)$, the probability of observing a value $E$ of the total energy. The symbols correspond to $M=4$ (pluses), $M=10$ (triangles), $M=20$ (asterisks), $M=25$ (squares), and $M=100$ (rhombi). These values have been obtained for $M\le 25$ by the highly precise numerical method detailed in the text, and for $M=100$ by a Monte Carlo simulation which can not accurately capture the tails of the distribution. In all cases we have set $T_1=1$ and $T_2=2$, corresponding to an effective temperature $T_\textrm{eff}=1.40473$, as obtained from Eq.~(\ref{teff}) (for $\epsilon=2,\,k=1$). The same data are plotted on a linear scale in panel (a) and on a logarithmic scale in panels (b) and (c). In panel (c) we have shifted the data of $M=20$ and $M=25$ downwards for the sake of clarity. Note that the effective form Eq.~(\ref{poeff}) provides an extremely good fit to the data (there are no free parameters), with increasing error  for larger values of the energy $E$.}
\end{figure}

It is clear that, as detailed balance is not satisfied, except for $T_1=T_2$, the stationary distribution can not be expressed as the canonical distribution $P_\textrm{st}(\{s\})\propto e^{-{\cal H}(\{s\})/kT}$. However, it is possible to find a canonical distribution at an effective temperature that provides a good approximation to the probabilities for the different energy values. We start by noting that, by using Eq.~(\ref{pstepsi})  the, exact, average energy of the chain is
\begin{equation}
\langle {\cal H}\rangle=\frac{\epsilon}{2}\sum_{i=1}^{M-1}\langle \ell_i\rangle=-\frac{\epsilon}{2}\frac{M-1}{2}\left[\tanh(\epsilon/2kT_1)+\tanh(\epsilon/2kT_2)\right].
\end{equation}
Comparing with the energy on an equilibrium chain at temperature $T$,  \makebox{$U=-\frac{\epsilon}{2}(M-1)\tanh(\epsilon/2kT)$} it is possible to define an effective (average) temperature as 
\begin{equation}\label{teff}
\tanh(\epsilon/2kT_\textrm{eff})=\frac{1}{2}\left[\tanh(\epsilon/2kT_1)+\tanh(\epsilon/2kT_2)\right].
\end{equation}
A one-dimensional Ising model with this effective temperature has the following expression for the (equilibrium) probability of an energy value $E$:
\begin{equation}\label{poeff}
p(E)=\frac{\Omega(E)e^{-E/Tk_\textrm{eff}}}{\sum_E \Omega(E)e^{-E/kT_\textrm{eff}}},
\end{equation}
being $\Omega(E)=\displaystyle 2{M\choose \frac{M-1-2E/\epsilon}{2}}$ the number of states with total energy ${\cal H}=\frac{\epsilon}{2}E$.  
In Fig.~\ref{fig:peff} we provide numerical evidence that this equilibrium distribution with the above introduced effective temperature provides a surprisingly good fit for the probability of having a total energy $E$. 

The fact that $p(E)$ can be approximated by an effective canonical distribution does of course not imply that the probability for a configuration has an equilibrium form. In summary, while
\begin{eqnarray}
P_\textrm{st}(\{s\})&\ne& {\cal Z}^{-1}e^{-{\cal H}(\{s\})/T_\textrm{eff}},
\end{eqnarray}
the probability for the total energy  is well approximated by 
\begin{eqnarray}
p(E)&=&\sum_{\{s\}|{\cal H}(\{s\})=E}P_\textrm{st}(\{s\})\approx {\cal Z}^{-1}\Omega(E)e^{-E/T_\textrm{eff}}.
\end{eqnarray}
Although we have checked that the canonical distribution is not exact for small values of the system sizes, we have no simple explanation for the goodness of this fit and we leave for further work the detailed analysis of its quality as a function of system size, temperature and other parameters of the model.

\bibliographystyle{unsrt}

\end{document}